\documentclass[prl,footinbib,superscriptaddress,reprint]{revtex4-2}
\pdfoutput=1
\usepackage{braket,dsfont,subfigure,amsfonts,amssymb,bm,graphicx,float,amsmath,hyperref,amsthm,txfonts}
\usepackage{orcidlink}
\usepackage{mathrsfs}
\usepackage{soul, color, xcolor}
\newcommand{\Tr}{{\rm Tr}}

\newcommand{\rrangle}{\rangle\!\rangle}

\newcommand{\llpipe}{|}
\newcommand{\sket}[1]{\ensuremath{\llpipe#1\rrangle}}
\usepackage[switch]{lineno}
\setstcolor{red}
\let\oldequation\equation
\let\oldendequation\endequation
\renewenvironment{equation}
{\linenomathNonumbers\oldequation}
{\oldendequation\endlinenomath}
\begin{document}
\title{Quantum State Tomography with Locally Purified Density Operators and Local Measurements}
\author{Yuchen Guo~\orcidlink{0000-0002-4901-2737}}
\affiliation{State Key Laboratory of Low Dimensional Quantum Physics and Department of Physics, Tsinghua University, Beijing 100084, China}
\author{Shuo Yang~\orcidlink{0000-0001-9733-8566}}
\email{shuoyang@tsinghua.edu.cn}
\affiliation{State Key Laboratory of Low Dimensional Quantum Physics and Department of Physics, Tsinghua University, Beijing 100084, China}
\affiliation{Frontier Science Center for Quantum Information, Beijing 100084, China}
\affiliation{Hefei National Laboratory, Hefei 230088, China}
\begin{abstract}
    Understanding quantum systems is of significant importance for assessing the performance of quantum hardware and software, as well as exploring quantum control and quantum sensing.
    An efficient representation of quantum states enables realizing quantum state tomography with minimal measurements.
    In this study, we propose an alternative approach to state tomography that uses tensor network representations of mixed states through locally purified density operators and employs a classical data postprocessing algorithm requiring only local measurements.
    Through numerical simulations of one-dimensional pure and mixed states and two-dimensional pure states up to size $8\times 8$, we demonstrate the efficiency, accuracy, and robustness of our proposed methods.
    Experiments on the IBM and Quafu Quantum platforms complement these numerical simulations.
    Our study opens avenues in quantum state tomography for two-dimensional systems using tensor network formalism.
\end{abstract}
\maketitle
\section{Introduction}
Quantum state tomography plays a fundamental role in characterizing and evaluating the quality of quantum states produced by quantum devices.  
It serves as a crucial element in the advancement of quantum hardware and software, regardless of the underlying physical implementation and potential applications~\cite{Nielsen2009, Preskill2018, Gebhart2023}.
However, reconstructing the full quantum state becomes prohibitively expensive for large-scale quantum systems that exhibit potential quantum advantages~\cite{Arute2019, Kim2023}, as the number of measurements required increases exponentially with system size.

Recent protocols try to solve this challenge through two main steps: efficient parameterization of quantum states and utilization of carefully designed measurement schemes and classical data postprocessing algorithms.
For one-dimensional (1D) systems with area law entanglement, the matrix product state (MPS)~\cite{Verstraete2006A, Perez2007, Verstraete2008, Schollwoeck2011, Orus2014, Cirac2021, Verstraete2023} provides a compressed representation.
It requires only a polynomial number of parameters that can be determined from local or global measurement results.
Two iterative algorithms using local measurements, singular value thresholding (SVT)~\cite{Cramer2010} and maximum likelihood (ML)~\cite{Baumgratz2013A}, have been demonstrated in trapped-ion quantum simulators with up to $14$ qubits~\cite{Lanyon2017}.
However, SVT is limited to pure states and thus impractical for noisy intermediate-scale quantum (NISQ) systems. 
Meanwhile, although ML can handle mixed states represented as matrix product operators (MPOs)~\cite{Pirvu2010, Jarkovsky2020}, it suffers from inefficient classical data postprocessing.
Another scheme reconstructs the quantum state by inverting local measurements, but the resulting MPO is not necessarily positive~\cite{Baumgratz2013B}.

On the other hand, some approaches are also based on tensor networks (TN) but update parameters from the sampling output of global measurements throughout the system~\cite{Han2018, Wang2020}.
It is noted that Torlai \textit{et al.}~\cite{Torlai2023} provides a process tomography method using global measurements, which can be applied directly to state tomography due to the Choi-Jamio\l{}kowski isomorphism~\cite{Jamiolkowski1972, Choi1975}.
However, these approaches cannot easily incorporate error mitigation techniques that focus mainly on estimators rather than samplers~\cite{Li2017, Temme2017, Endo2018, McArdle2019, Cai2021, Guo2022, Berg2023, Cai2023}, especially for readout errors~\cite{Berg2022, Yang2022}.
To scale to larger systems and integrate with error mitigation techniques, TN-based methods using only local measurements are preferred. 
However, TN state tomography remains unexplored for higher-dimensional quantum systems.

Besides tensor network state tomography, neural networks emerge as strong competitors for reconstructing unknown quantum states~\cite{Torlai2018, Carrasquilla2019, Torlai2020, Golubeva2022, Zhu2022, Zhao2024}, which can be expected from the deep connection between these two architectures~\cite{Chen2018, Huang2021, Wang2023}.
We note that all aforementioned methods presuppose that the target quantum state to be reconstructed can be efficiently approximated by a certain type of classical ansatz with only polynomial parameters,
Otherwise, reconstructing the entire density matrix, which possesses intrinsic exponential complexity, becomes infeasible~\cite{Zhao2023}.
For quantum states in the NISQ era, these requirements are generally satisfied, as circuit noise will inevitably limit the generation of entanglement, making it challenging for the resulting state to reach the regime of entanglement volume law~\cite{Noh2020}.
Another intriguing avenue is to estimate the observables without reconstructing the entire density matrix but directly from its classical shadows, which applies to a general quantum state.
Various techniques in this direction have been proposed for both quantum states~\cite{Aaronson2018, Huang2020, Hu2022, Liu2023} and quantum processes~\cite{Kunjummen2023, Levy2024}.

In this work, we introduce an approach to reconstructing mixed quantum states using only local expectation values, for which readout error mitigation is feasible.
We propose representing mixed states as locally purified density operators (LPDOs)~\cite{Verstraete2004A} and optimizing their parameters using a variant of gradient descent for a local loss function.
To validate the effectiveness of our method, we perform numerical simulations using typical quantum states: 1D critical Ising ground states, 1D gapless spin-$\frac{1}{2}$ Heisenberg ground states subjected to various noise models, two-dimensional (2D) random projected entangled pair states (PEPSs)~\cite{Verstraete2006B, Schuch2007, Perez2008}, 2D random isometric tensor network states (isoTNSs)~\cite{Zaletel2020, Soejima2020, Kadow2023, Liu2024}, and 2D cluster states.
Furthermore, we implement experiments on real quantum devices accessible through the IBM and Quafu Quantum platforms.

\section{Results}
\subsection{Quantum state tomography with LPDO and local loss function}
In this section, we introduce our reconstruction approach for tomography of mixed states. 
We begin by efficiently parameterizing mixed states using LPDOs, which have been shown to be efficient in studying thermal or dissipative many-body systems in 1D~\cite{Werner2016, Cheng2021, Guo2023C, Guo2024}.

In the LPDO form, a 1D mixed quantum state is expressed as follows
\begin{equation}
        \hat{\rho} = \sum_{\{\bm{\tau}, \bm{\omega}\}}\sum_{\{\bm{\mu}, \bm{\nu}, \bm{\kappa}\}}\prod_{j=1}^N[A_j]^{\tau_j, \kappa_{j}}_{\mu_{j-1}, \mu_{j}}[A_j^*]^{\omega_j, \kappa_{j}}_{\nu_{j-1}, \nu_{j}}\ket{\tau_1,\cdots,\tau_N}\hspace{-1mm}\bra{\omega_1,\cdots,\omega_N},
\end{equation}
where $\bm{\mu}$ and $\bm{\nu}$ denote virtual indices that describe quantum entanglement, and $\bm{\kappa}$ represents inner indices (also known as Kraus indices) introduced for open systems~\cite{Cheng2021}, as depicted in Fig.~\ref{Fig: Compare}(a).
Importantly, an LPDO is constructed to be Hermitian and positive semidefinite by design.
In scenarios with weak local noise, the Kraus dimension $d_{\kappa}$ is typically a small constant, independent of the system size $N$.
This enables a significantly more efficient tomography approach compared to directly reconstructing an MPO.
Note that the recently proposed quantum process tomography method~\cite{Torlai2023} has adopted the LPDO variant to represent quantum channels.
In addition, the ML method~\cite{Baumgratz2013A} could be adapted and integrated into this framework by replacing MPO iterations with computationally less expensive LPDO operations.
However, we do not explore this further as the exponential computational cost of iteration (shown later) is not easily reduced.

Next, we propose a loss function constructed from local measurements and a variant of gradient descent to update the local LPDO tensors.
The loss function for our problem is chosen as
\begin{equation}
    \Theta = \sum_{i}{||\hat{\sigma}_{\braket{i}} - \hat{\rho}_{\braket{i}}||_F^2} \equiv \sum_{i} \Theta_{\braket{i}},\label{equ: Cost}
\end{equation}
where $\hat{\sigma}_{\braket{i}}$ and $\hat{\rho}_{\braket{i}}$ are the reduced density matrices for the sites $\{i, \cdots, i+L-1\}$ of the target state obtained from experiments (see Methods) and the reconstructed state respectively, as shown in Fig.~\ref{Fig: Compare}(a). 
Expanding each term in the loss function gives
\begin{equation}
    \Theta_{\braket{i}} = \Tr{\left[\hat{\sigma}_{\braket{i}}^2 - 2\hat{\sigma}_{\braket{i}}\hat{\rho}_{\braket{i}}+\hat{\rho}_{\braket{i}}^2\right]}.
\end{equation}
To calculate the gradients, terms such as
\begin{equation}
    \frac{\partial \Theta_{\braket{i}}}{\partial A_j^*} = 2 \Tr{\left[\left(\hat{\rho}_{\braket{i}}-\hat{\sigma}_{\braket{i}}\right)\frac{\partial\hat{\rho}_{\braket{i}}}{\partial A_j^*}\right]},
\end{equation}
need to be computed, requiring $O(N^2)$ computational complexity.

However, instead of directly optimizing the entire loss function, we update each local tensor $A_j$ only considering the adjacent terms $\Theta_{\langle i \rangle}$ that involve the target site $j$ in the loss function.
Specifically, we update $A_j$ according to the following rule
\begin{equation}
    A_{j} \rightarrow A_{j} - \eta \sum_{i=j-L+1}^{j}\frac{\partial \Theta_{\braket{i}}}{\partial A_{j}^{*}},\label{equ: Gradient}
\end{equation}
where $\eta$ is the learning rate, automatically adjusted using the Adam optimizer~\cite{Kingma2017}.
Evaluating the gradient in each iterative step has a time complexity of $O(ND^3)$, where $D$ is the virtual bond dimension. 
This approach converges to a high-quality approximation of the target state in only $O(\log(N))$ iterative steps without encountering the issue of local minima or barren plateaus, as demonstrated below.
LPDO and the local loss function together constitute our Grad-LPDO method.
Importantly, for pure states, one can set $d_{\kappa}=1$ and optimize the MPS manifold using the same gradient method, which is treated as a special version of Grad-LPDO.

Here we briefly outline the relationship between our approach and previous tensor network state tomography methods, which generally fall into two categories.
Our method follows the category of optimizing a local loss function, encoding only local properties estimated from the measurements for local observables~\cite{Cramer2010, Baumgratz2013A, Lanyon2017}.
The analytical theorem behind this scheme was previously discussed~\cite{Cramer2010} (also known as ground-state witness), i.e., for two pure quantum states $\ket{\phi}$ (target) and $\ket{\psi}$ (reconstructed), if $\ket{\psi}$ is the ground state of a local Hamiltonian $H=\sum_{i}{h_{\braket{i}}}$ with an energy gap $\Delta$, then the fidelity between these two states satisfies that
\begin{equation}
    f\geq 1-\frac{\Tr{\left[\sum_i{\hat{h}_{\braket{i}}\left(\hat{\sigma}_{\braket{i}}-\hat{\rho}_{\braket{i}}\right)}\right]}}{\Delta}.\label{equ: witness}
\end{equation}
In other words, for a quantum circuit with a constant depth, whose output state can be represented by an injective MPS that has a local, gapped parent Hamiltonian~\cite{Perez2007}, only local information is sufficient to reconstruct the entire state.
For a general mixed state, though there is no rigorous proof, we believe that such an argument also holds if the mixed state is generated from a noisy quantum circuit, where the mixed state can be regarded as a pure state perturbed by noise.
The alternative approach regards the tensor network as a generative model, using a global loss function to characterize the probability distributions of samples generated by simultaneous measurements on all qubits.
These measurements involve random bases, and each sample is represented by an integer-valued string that describes the measurement basis and spin configuration~\cite{Han2018, Wang2020, Torlai2023}.
The concrete realization and requirements of these two formalisms will be further compared and discussed later.

\subsection{Numerical simulations in different systems}
\subsubsection{Comparison between Grad-LPDO and ML-MPS}
\begin{figure*}
    \includegraphics[width=\textwidth]{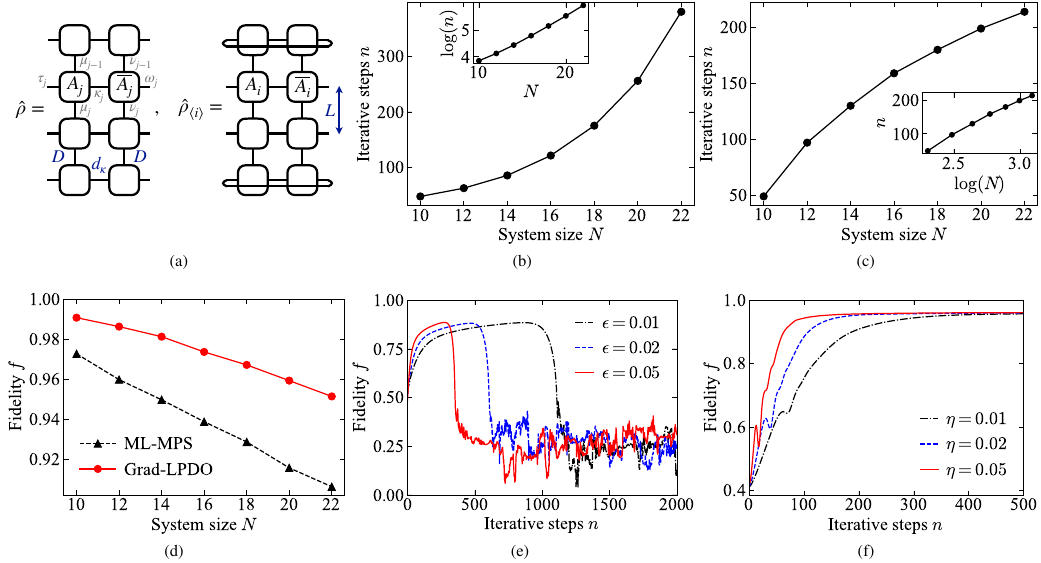}
    \caption{Schematic of Grad-LPDO method and performance on 1D critical Ising ground state.
    (a) Mixed state with $N=4$ in the locally purified density operator (LPDO) form and the reduced density matrix for the sites $\{i, i+1\}$.
    (b)-(f) Comparison of the ML-MPS and Grad-LPDO methods with bond dimension $D = D_0 = 16$ and varying system sizes $N$.
    (b)-(c) Number of steps $n$ to achieve fidelity $f=0.9$ for (b) ML-MPS with $\log(n)$ inset, and (c) Grad-LPDO with $\log(N)$ inset.
    (d) Maximal fidelity during iterations for both methods.
    (e)-(f) Convergence stability of two methods for different learning and update rates.
    (e) The ML-MPS method, where $\epsilon$ is the update rate defined under Eq.~\eqref{equ: ML-Fixed}.
    (f) Our Grad-LPDO method, where $\eta$ is the learning rate in the Adam optimizer.}
    \label{Fig: Compare}
\end{figure*}

In this section, we present numerical demonstrations of our Grad-LPDO method for both pure and mixed states of special interest.
We begin by comparing the performance of a standard approach also using local measurements, namely ML-MPS~\cite{Baumgratz2013A}, and our proposed Grad-LPDO method for reconstructing the ground state of the 1D transverse field Ising model at the critical point $g=1$ under open boundary condition (OBC).
Both methods share a common initial step for calculating local reduced density matrices from experimental results for subsequent construction.
Hence, in terms of experimental complexity, these two methods are identical.
The critical comparison therefore lies in the complexity of classical data postprocessing algorithms.

The Hamiltonian is given by
\begin{equation}
    H = - \sum_{\braket{i, j}} Z_iZ_j+g\sum_i X_i.
\end{equation}
The ground state is approximated by an MPS with $D_0 = 16$ using the standard variational method~\cite{Crosswhite2008, Orus2014}, which serves as the target state for subsequent reconstruction.
Here, we assume that all measurements are ideal and directly construct the reduced density matrices $\hat{\sigma}_{\braket{i}}$ from the target state.
Thus, any reconstruction errors are solely attributed to insufficient local measurements and inefficient classical data postprocessing.
We perform both algorithms for different system sizes $N$, with $D = D_0 = 16$ and $L = 2$, where the initial states are chosen as the paramagnetic ground state for $g\rightarrow+\infty$.
The hyperparameters in the Adam optimizer are set as $\xi_1 = \xi_2 = 0.8$ and $\epsilon = 10^{-8}$.
Fig.~\ref{Fig: Compare}(b)-(c) shows the number of iterative steps required to achieve fidelity $f = 0.9$ for different system sizes, representing the convergence speed of the algorithms.

The inset of Fig.~\ref{Fig: Compare}(b) clearly shows that the time complexity of the ML-MPS method scales exponentially with system size for a given reconstruction accuracy.
In contrast, the results shown in Fig.~\ref{Fig: Compare}(c) are highly promising, indicating that the number of gradient steps required for convergence scales only as $O(\log{N})$.
In other words, the overall time complexity of our Grad-LPDO method to reconstruct a pure 1D state is $O(N\log(N)D^3)$.
This time complexity significantly outperforms ML-MPS (and SVT-MPS with time complexity $O(N^4D^4)$~\cite{Cramer2010}), meaning that our method saves considerable time compared to ML-MPS, especially for large systems.
Even for small systems, ML-MPS holds no efficiency advantage since truncating MPS in each iteration, as required in the ML method (see Methods), is generally more computationally expensive than simply contracting the environment in Grad-LPDO.

To provide a comprehensive comparison of these two methods, we calculate the maximal fidelity achieved during iterations for different $N$ in Fig.~\ref{Fig: Compare}(d), which demonstrates the higher accuracy of our method.
At the same time, we also consider convergence stability when comparing different methods. 
Typical iteration curves plotted in Fig.~\ref{Fig: Compare}(e)-(f) reveal that Grad-LPDO consistently converges to maximal fidelity, while the convergence of ML-MPS is less stable, regardless of the update or learning rate per step.
In conclusion, our Grad-LPDO method surpasses the previous ML-MPS method in three key aspects: efficiency, accuracy, and stability.

\begin{figure*}
    \includegraphics[width=\textwidth]{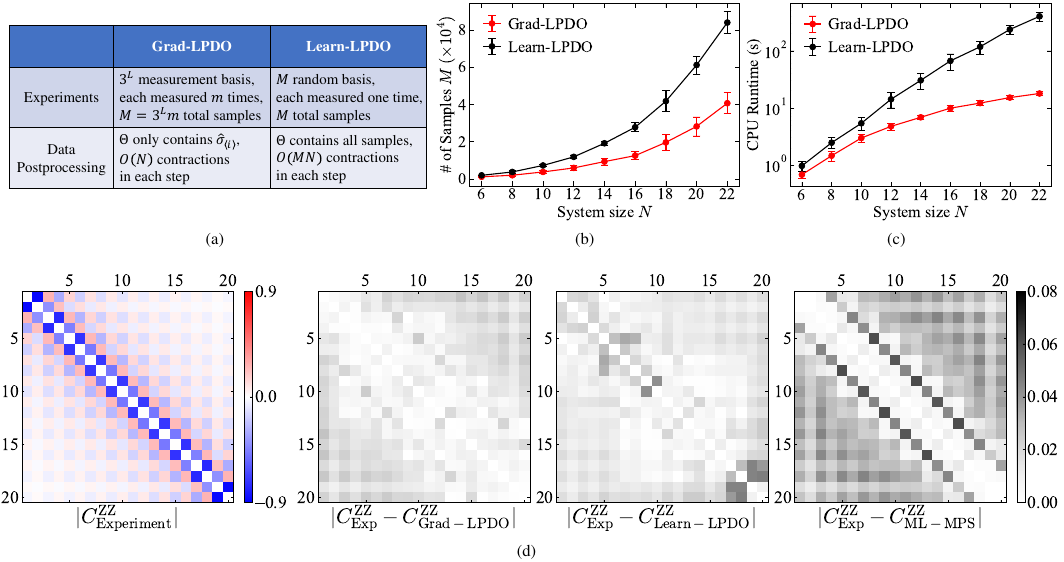}
    \caption{Comparison of different methods for one-dimensional spin-$1/2$ Heisenberg ground states.
    (a)-(c) Comparison of Grad-LPDO and Learn-LPDO.
    (a) Different schemes for experiments and data postprocessing of two methods.
    (b) Number of samples to achieve fidelity $f=0.95$ of two methods.
    (c) Classical runtime on a single laptop CPU to achieve $f=0.95$ of two methods.
    (d) ZZ correlation function for the target state and differences in the correlation between the target state and the reconstructed states for three methods with system size $N=20$ and bond dimension $D=D_0=16$.
    All results are averaged over $100$ runs to simulate statistical measurement errors, along with the standard deviation.}
    \label{Fig: Heisenberg}
\end{figure*}

\subsubsection{Comparison between Grad-LPDO and Learn-LPDO}
Proceeding to the comparison with a recent process tomography method using global measurements, referred to as Learn-LPDO~\cite{Torlai2023}, we illustrate the experimental settings of these two formalisms in Fig.~\ref{Fig: Heisenberg}(a).
In Learn-LPDO, each sample is generated from the circuit on a random measurement basis in a single shot.
On the other hand, in our method (as well as other local information-based approaches~\cite{Cramer2010, Baumgratz2013A, Lanyon2017}), each reduced density matrix of length $L$ must be estimated from the complete local bases of size $4^L-1$, providing $(N-L+1)(4^L-1)$ observables to be measured.
Fortunately, not all these measurements are independent, and the total number of independent bases can be reduced to only $3^L$, regardless of the system size $N$ (see Methods).
If each basis is measured $m$ times to obtain an accurate expectation value, the total number of experimental samples is $M = 3^L m$.

In the following, we examine the performance of these two methods on the ground state of the 1D gapless spin-$\frac{1}{2}$ Heisenberg model
\begin{equation}
    H = \sum_{\braket{i, j}}\mathbf{S}_{i}\cdot\mathbf{S}_{j}
\end{equation}
with $N=20$ and $D_0=16$.
With such a model exhibiting non-trivial long-range antiferromagnetic correlation, one might expect that the global Learn-LPDO method would perform better than our Grad-LPDO method, since the former takes into account global measurements while the latter depends only on local marginals.
However, the following results will show that this is not the case.

Fig.~\ref{Fig: Heisenberg}(b) shows the number of total samples $M$ required for a given reconstruction accuracy of $f = 0.95$ for each method.
Here, both sampling processes are simulated, unlike in the previous section, to include statistical errors from measurements.
The result reveals that our Grad-LPDO method with $L=2$ has a similar scaling behavior of experimental complexity compared to Learn-LPDO, with a constant coefficient difference of approximately $\sim 1/2$.
In particular, even with the same number of samples, Learn-LPDO would be more difficult to realize than Grad-LPDO, since all samples in Learn-LPDO are generated from random bases that differ from each other, while the measurement settings are much simpler in Grad-LPDO, involving only $3^L$ different configurations.
Therefore, our method proves to be slightly more efficient than Learn-LPDO in terms of experimental complexity.

Regarding the classical data postprocessing complexity of the reconstruction process, we evaluate the CPU runtime required to converge to a fidelity of $f=0.95$ serving as a figure of merit.
Both methods are implemented on a laptop CPU (Intel i7-11800H).
The results shown in Fig.~\ref{Fig: Heisenberg}(c) demonstrate a substantial advantage of orders of magnitude for our Grad-LPDO method, where the gap further increases with the system size $N$.
The significant difference in the classical runtime can be attributed to the complexity of evaluating gradients in each iterative step.
As discussed in the previous section, our Grad-LPDO method uses samples from experiments to construct reduced density matrices $\hat{\sigma}_{\braket{i}}$, then the loss function $\Theta$ in Eq.~\eqref{equ: Cost} involves only $N-L+1$ terms of $\hat{\sigma}_{\braket{i}}$ instead of the original samples.
In essence, Grad-LPDO requires the contraction of $NL$ tensor networks when estimating all the gradients in Eq.~\eqref{equ: Gradient}.
In contrast, Learn-LPDO directly incorporates all samples into the loss function, requiring the contraction of $MN$ tensor networks, as each sample contributes one tensor network to one gradient.
In summary, to achieve a comparable reconstruction accuracy, the Learn-LPDO method proves to be more costly both experimentally and computationally than our Grad-LPDO method.
Meanwhile, in our method, each complexity increases polynomially with system size $N$ rather than exponentially, indicating the scalability of our method for gapless states in larger systems.

To gain an intuitive understanding of the performance of the above three methods, the ZZ correlation function $C^{\rm{ZZ}}_{ij} = \braket{Z_iZ_j}-\braket{Z_i}\braket{Z_j}$ is visualized for the target state in Fig.~\ref{Fig: Heisenberg}(d).
We also show the differences in the correlation between the target state and the reconstructed states with $D=D_0=16$.
The result indicates that Grad-LPDO can capture most of the long-range correlation and antiferromagnetic order from only local measurements.
On the other hand, ML-MPS cannot faithfully reproduce the correlation between sites with intervals $L\geq 3$, which limits its application to systems with non-trivial orders.
For the Learn-LPDO method, the overall accuracy is similar to that of Grad-LPDO, but with a much larger fluctuation in the correlation function due to the intrinsic limitations of any learning-based method.

\subsubsection{Numerical simulations for 1D mixed states}
\begin{figure*}
    \includegraphics[width=\textwidth]{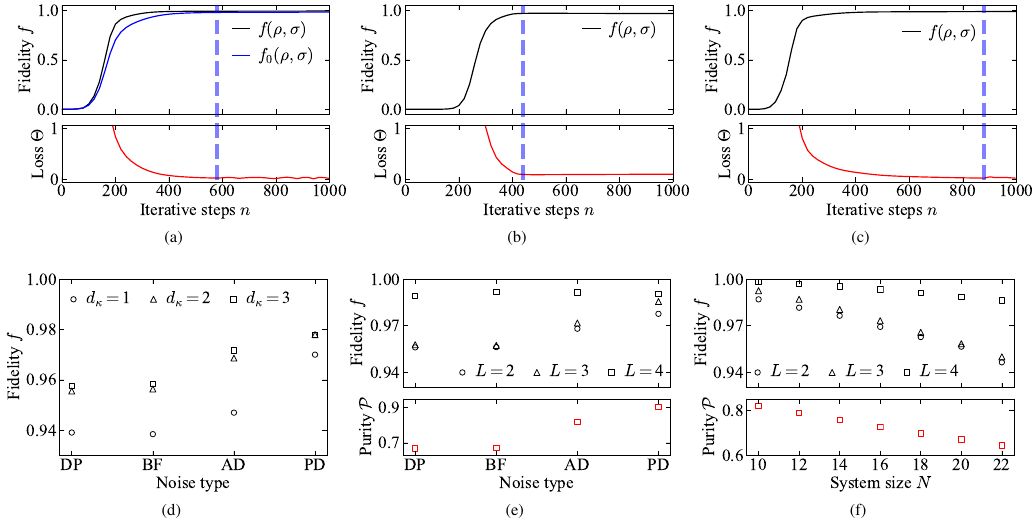}
    \caption{Results for gapless spin-$\frac{1}{2}$ Heisenberg ground states without noise or undergoing various types of noise.
    (a-c) Conventional fidelity $f_0$, our alternative fidelity $f$, and the loss function $\Theta$ during the iteration process for bond dimension $D=D_0=16$. Blue dashed lines show the consistent convergence over different indicators.
    (a) $N = 10$ spins with amplitude damping (AD) noise with error rate $\varepsilon=0.01$, reconstructed by locally purified density operators (LPDOs) with local measurement range $L=2$ and Kraus dimension $d_{\kappa}=2$.
    (b) $N = 20$ spins without noise, reconstructed by LPDOs with $L=2$ and $d_{\kappa}=1$.
    (c) $N = 20$ spins with bit flipping (BF) noise with $\varepsilon=0.01$, reconstructed by LPDOs with $L=2$ and $d_{\kappa}=2$.
    (d-f) Maximal fidelity $f$ and purity $\mathcal{P}$ for different system sizes $N$ and various types of noise, where purity $\mathcal{P}$ is irrelevant to $L$.
    (d) $N = 20$ spins with four types of noise with $\varepsilon=0.01$, including depolarizing (DP), BF, AD, and phase damping (PD) noise, reconstructed by LPDOs with $L=2$ and different $d_{\kappa}$.
    (e) $N = 20$ spins with four types of noise with $\varepsilon=0.01$, reconstructed by LPDOs with $d_{\kappa}=2$ and different $L$.
    (f) DP noise with $\varepsilon=0.01$ added to systems with different $N$, reconstructed by LPDOs with $d_{\kappa}=2$ and different $L$. }
    \label{Fig: Mixed}
\end{figure*}

Next, we consider Heisenberg ground states with $D_0=16$ that undergo different types of local noise.
Specifically, we add four types of single-qubit noise with an equal error rate $\varepsilon=0.01$ to each qubit of the ideal state, including depolarizing (DP), bit flipping (BF), amplitude damping (AD), and phase damping (PD).
Fig.~\ref{Fig: Mixed} shows numerical results, where we reconstruct target states using LPDOs with $D=16$ and different $d_{\kappa}$, along with accurate local measurements of length $L=2$ without statistical errors.
The fidelity between two density matrices is defined as their inner product in the operator space $f\left(\hat{\rho}_1, \hat{\rho}_2\right) \equiv \Tr{\left(\hat{\rho}_1\hat{\rho}_2\right)} / \sqrt{\Tr{\left(\hat{\rho}_1^2\right)}\Tr{\left(\hat{\rho}_2^2\right)}}$, which can be efficiently calculated for LPDOs (see Methods).
To clarify the relationship between different figures of merits, we compare the conventional definition of fidelity $f_0\left(\hat{\rho}_1, \hat{\rho}_2\right) = \left(\Tr{\sqrt{\sqrt{\hat{\rho}_1}\hat{\rho}_2\sqrt{\hat{\rho}_1}}}\right)^2$ (which is intractable for large systems), our alternative definition $f$, and the loss function $\Theta$ throughout the iteration process in Fig.~\ref{Fig: Mixed}(a)-(c) for different system sizes and with or without noise.
The results show the consistent convergence (blue dashed lines) of these indicators in different situations.
This not only confirms the validity of our alternative definition $f$ but also highlights the sensitivity of the loss function $\Theta$ as a reliable convergence criterion, especially considering the impracticality of evaluating fidelity in real experiments.

Introducing noise to the target state will generally reduce the reconstruction fidelity $f$, which cannot be improved by only increasing $d_{\kappa}$ as implied in Fig.~\ref{Fig: Mixed}(d).
This decrease arises because local measurements alone cannot distinguish between mixtures in the reduced density matrix from entanglement with unmeasured qubits (represented by virtual indices) and from quantum noise (represented by Kraus indices), as confirmed by our numerical simulations.
For example, reconstructing quantum states with depolarizing or bit-flipping noise poses greater challenges, since these noise models are stochastic and directly introduce mixtures across different trajectories.
Furthermore, the purity of the target state $\mathcal{P} = \Tr[\hat{\rho}_0^2]$ is plotted in Fig.~\ref{Fig: Mixed}(e) for different noise types and in Fig.~\ref{Fig: Mixed}(f) for depolarizing noise across varying system sizes $N$.
They show a clear dependence of the reconstruction fidelity $f$ on $\mathcal{P}$, supporting our earlier arguments.

To overcome this challenge, we increase the measurement length $L$ in Fig.~\ref{Fig: Mixed}(e)-(f), significantly improving the accuracy for all types of noise.
With adjacent $4$-site measurements, the reconstruction fidelity $f$ exceeds 0.985 even for the most challenging depolarizing noise with up to $N=20$ qubits.
However, the experimental and data postprocessing costs scale exponentially with $L$, requiring a trade-off between accuracy and efficiency.
Our results indicate that even for moderate-sized critical systems with typical noise levels, a small constant $L$ suffices for high-fidelity reconstruction at an acceptable cost using modern experimental and numerical techniques.

\subsubsection{Generalization to 2D systems}
\begin{figure*}
    \includegraphics[width=\linewidth]{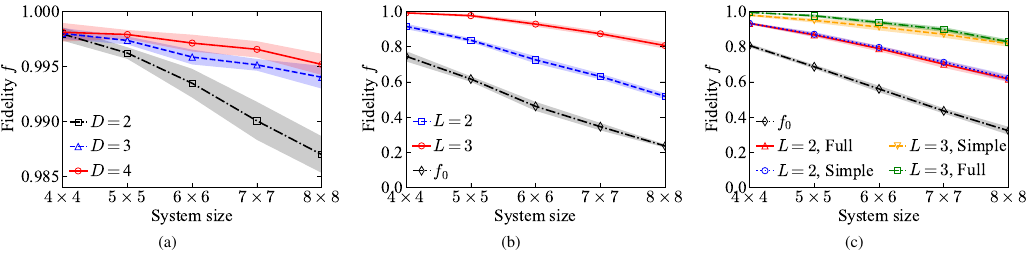}
    \caption{
    Generalization of Grad-LPDO to 2D projected entangled pair state (PEPS).
    (a) Reconstruction fidelity $f$ for random PEPSs with target bond dimension $D_0 = 3$ and different reconstructed bond dimensions $D$, with local measurements of size $1\times 2$ and $2\times 1$ and simple gradient.
    (b) Reconstruction fidelity $f$ for noisy random isoTNSs with $D = D_0 = 2$ and varying system sizes, with local measurements of size $1\times L$ and $L\times 1$ and simple gradients, compared to the initial fidelity $f_0$ between ideal and noisy states.
    (c) Reconstruction fidelity $f$ for noisy cluster states with $D = D_0 = 2$ and varying system sizes, with local measurements of size $1\times L$ and $L\times 1$ and different strategies to calculate gradients (simple or full).
    All results are averaged over $100$ realizations of the target states along with the standard deviation.
}
    \label{Fig: PEPS}
\end{figure*}

We now generalize our Grad-LPDO method to pure states of two-dimensional systems on square lattices, represented as PEPSs.
The key idea and procedure remain the same, where the local measurements in Eq.~\eqref{equ: Cost} are applied to the $L_1\times L_2$ subsystems.
When updating the corresponding local tensor $A_j$, gradients in Eq.~\eqref{equ: Gradient} contain terms with local measurements covering the site $j$, which we denote as simple gradients.
In general, the contraction of a 2D TN is computationally expensive.
Here, we adopt the standard truncation method for finite-size systems~\cite{verstraete2004B, Orus2014}.
The 2D TN with bond dimension $D$ is contracted layer by layer as the evolution of a 1D MPS, truncating the bond dimension to $\chi=D^2$ after each contraction.

We first simulate random PEPS (i.e., all tensor elements are random complex numbers) as target states with fixed $D_0=3$ and varying system sizes $N\times N$, where random PEPSs with different $D$ are chosen as initial states for the iteration.
We directly construct $\hat{\sigma}_{\braket{i}}$ on all $1\times 2$ and $2\times 1$ subsystems of the target states.
Fig.~\ref{Fig: PEPS}(a) shows the reconstruction fidelity of the $100$ shots for each pair of $N$ and $D$, with error bars giving the standard deviation.
The average fidelity reaches $0.995$ for random $8\times8$ PEPSs with $D_0 = 3$, which are generally noncritical.

However, a PEPS with random tensor elements is generally slightly entangled and thus does not represent a good 2D generalization.
Therefore, we proceed to two more specific types of 2D system, including the isometric tensor network (isoTNS)~\cite{Zaletel2020} and the cluster state.
The former has found wide applications in topological states~\cite{Soejima2020, Liu2024}, quantum circuits~\cite{Slattery2021, Lin2022}, and thermal states~\cite{Kadow2023}, while the latter is a typical state with symmetry protected topological order~\cite{Gu2009, Chen2010, Pollmann2010, Chen2011, Pollmann2012, Chen2013} and serves as a resource state for measurement-based quantum computation~\cite{Briegel2001, Raussendorf2001, Briegel2009, Raussendorf2017, Raussendorf2019} or measurement-induced phase transitions~\cite{Liu2022, Guo2023B}.
As in real experiments for benchmarking quantum devices or error mitigation, one knows the ideal quantum state representation and tries to learn the noisy quantum state, we can choose the ideal state as the initial ansatz for iteration to speed up the convergence.
To be more specific, we first construct an ideal PEPS $\ket{\Psi_{\rm Ideal}}$ and add perturbation to each local tensor as an effective noise model, where the output state $\ket{\Psi_{\rm Target}} = \ket{\Psi_{\rm Noisy}}$ is set as the target state for reconstruction.
Meanwhile, the ideal state is chosen as the initial state for iteration $\ket{\Psi_0} = \ket{\Psi_{\rm Ideal}}$.

In Fig.~\ref{Fig: PEPS}(b), we simulate our tomography method on random isoTNSs, where each local tensor corresponds to a random haar unitary.
In other words, the entire state can be mapped to a sequential quantum circuit with random quantum gates~\cite{Garnerone2010, Slattery2021}, whose depth will scale linearly with the system size given only a constant bond dimension $D$, and therefore has significant expressive power for complex quantum systems and serves as a strong candidate for benchmarking our approach.
Reconstruction fidelities with different local measurement sizes $1\times L$ and $L\times 1$ for $D=D_0=2$ and varying system sizes are compared, where the initial fidelity $f_0\equiv f(\ket{\Psi_{\rm Target}}, \ket{\Psi_{\rm 0}}) = f(\ket{\Psi_{\rm Noisy}}, \ket{\Psi_{\rm Ideal}})$ is plotted for benchmarking the `noise strength'.
The result demonstrates the effectiveness of our method for typical 2D systems up to $8 \times 8$ of interest to capture most of the noise effects with only local measurements of size $1\times 3$ and $3\times 1$.
In addition, we also compare the simple gradients and the full gradients (including all terms in the loss function~\eqref{equ: Cost}) in Fig.~\ref{Fig: PEPS}(c) for 2D cluster states, where full gradients are slightly more accurate than simple gradients, indicating the effectiveness of focusing only on local neighbors when implementing classical data postprocessing.

\subsection{Real experiments with cloud quantum computation}
\subsubsection{Experiments on IBM Quantum platform}
\begin{figure*}
    \includegraphics[width=\textwidth]{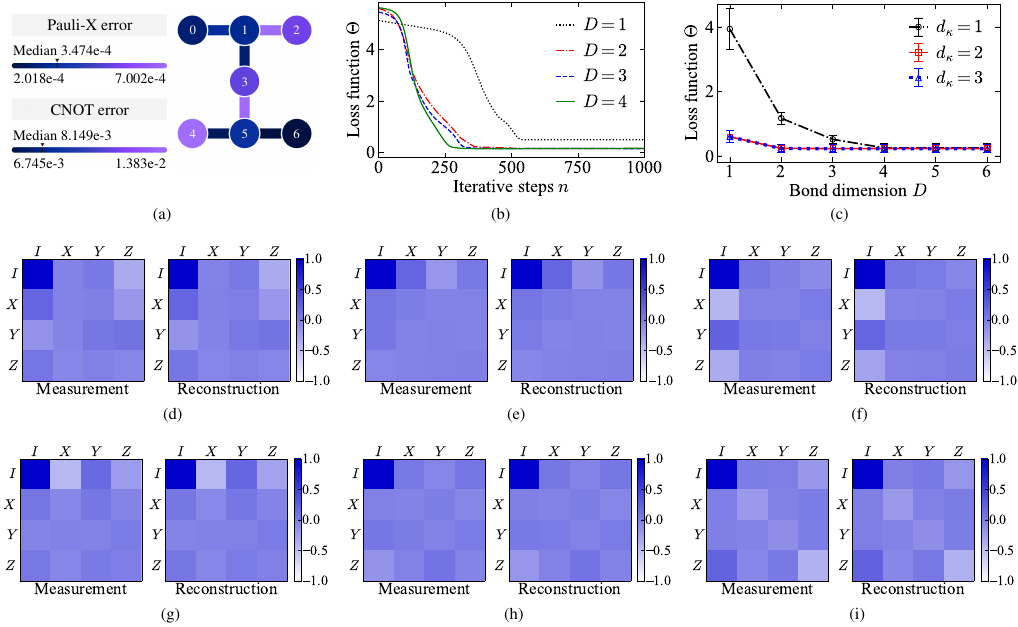}
    \caption{Experiments on IBM Quantum platform.
    (a) Configuration of qubits and noise information for device `ibm\_nairobi'.
    (b) The iteration process of one circuit realization ($\textrm{seed}=57$) for Kraus dimension $d_{\kappa}=2$ and different bond dimensions $D$.
    (c) Average residual loss $\Theta$ over $64$ random circuit realizations for different $D$ and $d_{\kappa}$ along with the standard deviation.
    (d)-(i) Expectation values for $L=2$ local observables obtained from experiments and reconstruction for $d_{\kappa}=2$ and $D=4$.
    (d) qubit (0, 1).
    (e) qubit (1, 2).
    (f) qubit (2, 3).
    (g) qubit (3, 4).
    (h) qubit (4, 5).
    (i) qubit (5, 6).}
    \label{Fig: IBM}
\end{figure*}

To demonstrate our method on real quantum hardware, we conduct experiments on the IBM Quantum platform.
Specifically, we use the `ibm\_nairobi' quantum computer, a superconducting processor with $N=7$ available qubits, whose qubit configuration and noise information are shown in Fig.~\ref{Fig: IBM}(a).
The input state is the trivial product state $\ket{\psi_0} = \ket{0}^{\otimes N}$, then random Haar circuits are implemented to generate the output states to be reconstructed.
$3^L$ ($L=2$ here) numbers of Pauli strings are measured to estimate each reduced density matrix $\hat{\sigma}_i$ for the target state, and each observable is measured using $m=10,000$ shots.
To mitigate errors in measurement results, the twirled readout error extinction (T-REx)~\cite{Berg2022} is employed.
Other error mitigation techniques for circuit errors are not considered, as one of the main objectives of tomography is to learn about the noise information in the system.
Since the target state is unknown in practice, we use the loss function $\Theta$ defined in Eq.~\eqref{equ: Cost} to represent the residual error for tomography, which is highly related to the final reconstruction fidelity that cannot be directly evaluated in practical scenarios~\cite{Cramer2010}.

In Fig.~\ref{Fig: IBM}(b), the iteration process of one typical circuit realization (seed $57$ for the random circuit) is plotted for $d_{\kappa}=2$ and different $D$.
The experimentally measured expectation values $\Tr{[\hat{\sigma}_{\braket{i}}\hat{P}_{\braket{i}}]}$ and the corresponding estimated values from the reconstructed states $\Tr{[\hat{\rho}_{\braket{i}}\hat{P}_{\braket{i}}]}$ are calculated for each reduced density matrix with $d_{\kappa}=2$ and $D=4$, where $\hat{P}_{\braket{i}}$ refers to Pauli strings.
These results are visualized in Fig.~\ref{Fig: IBM}(d)-(i) to demonstrate their high consistency.
Fig.~\ref{Fig: IBM}(c) shows the residual error averaged over $64$ random circuit realizations along with the standard deviation.
For each circuit realization, we choose $100$ random LPDOs as initial states and record the average residual error.
These results confirm that LPDOs with small $d_{\kappa}$ and $D$ serve as good approximations for quantum states generated from noisy quantum circuits, validating the performance of our tomography scheme.

\subsubsection{Experiments on Quafu Quantum platform}
\begin{figure*}
    \includegraphics[width=\textwidth]{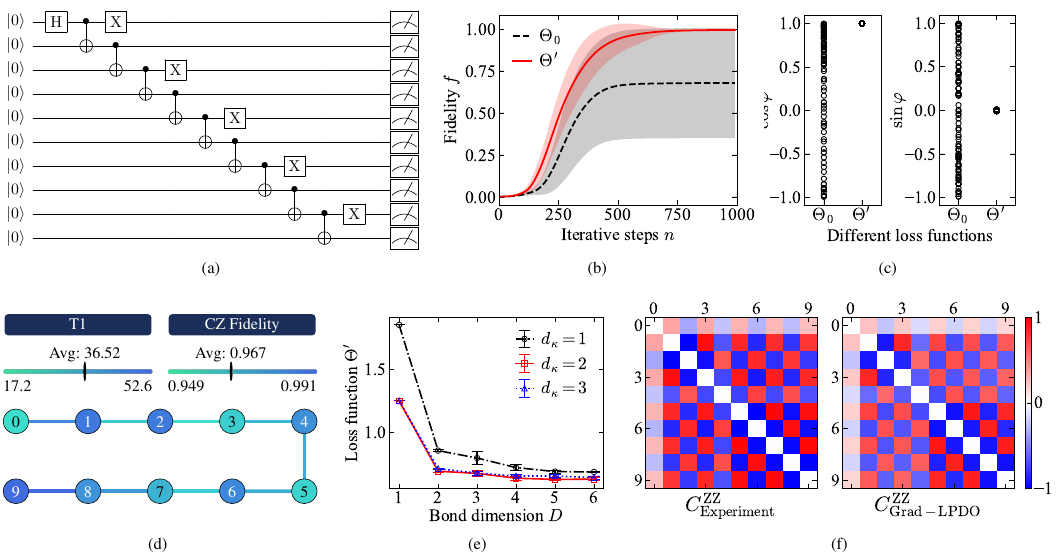}
    \caption{Experiments on Quafu Quantum platform.
    (a) Circuit configuration to generate an antiferromagnetic GHZ state with system size $N=10$.
    (b) Fidelity during the iteration process for the original loss function $\Theta$ and the modified $\Theta^{\prime}$.
    Both are averaged over $100$ runs of simulation along with the standard deviation.
    (c) Relative phase of two macroscopic components in the GHZ state for two loss functions.
    (d) Configuration of qubits and noise information for device `ScQ-P10'.
    (e) Averaged residual loss $\Theta$ for different bond dimensions $D$ and Kraus dimensions $d_{\kappa}$ over $64$ runs of experiments along with the standard deviation.
    (f) Values of the ZZ correlation function for the target state and the reconstructed state with $D=4$ and $d_{\kappa}=2$.}
    \label{Fig: Quafu}
\end{figure*}

Besides random quantum circuits, we also apply our approach to test on a prominent quantum state - the Greenberger-Horne-Zeilinger (GHZ) state characterized by long-range entanglement, which is of central importance in both quantum information and condensed matter physics.
The quantum circuit designed to generate a GHZ state is shown in Fig.~\ref{Fig: Quafu}(a), where we add extra $X$ gates on even qubits to create an antiferromagnetic order, which improves clarity in the subsequent visualization of the correlation function, resulting in the following target state
\begin{align}
    \ket{\Psi_0}=\frac{1}{\sqrt{2}}\left[\ket{\downarrow\uparrow\cdots\downarrow\uparrow}+\ket{\uparrow\downarrow\cdots\uparrow\downarrow}\right].
\end{align}
We first numerically simulate our tomography process for the ideal GHZ state with $N=10$, $D=D_0=2$, and $L=2$.
The iteration process is shown in Fig.~\ref{Fig: Quafu}(b), where the converged fidelity $f$ exhibits a random pattern over a wide range, arising from the degeneracy of the GHZ state that results in infinitely many states with the same reduced density matrices, i.e.,
\begin{align}
    \ket{\Psi_{\varphi}}=\frac{1}{\sqrt{2}}\left[\ket{\downarrow\uparrow\cdots\downarrow\uparrow}+e^{i\varphi}\ket{\uparrow\downarrow\cdots\uparrow\downarrow}\right],
\end{align}
with our target state corresponding to $\varphi=0$.
To distinguish between these states, we must determine the relative phase $\varphi$ of two macroscopic components, necessitating two global measurements of all qubits to evaluate the real and imaginary part of $e^{i\varphi}$ (the choice is not unique)
\begin{align}
    &\braket{\prod_{i=1}^N X_i} = \cos{\varphi},\\
    &\braket{\prod_{i=1}^{N/2} X_{2i-1}Y_{2i}} = \sin{\varphi},
\end{align}
for $N=10$.
Here we select these observables since the measurement configurations already exist in the $3^L$ basis used for calculating local reduced density matrices (see Methods).
In other words, no additional measurements are required for these two global observables.
The new loss function reads as
\begin{align}
    \Theta^{\prime} = \Theta_0 + \sum_{k}\left|\braket{O_k}-o_k\right|^2,
\end{align}
where $\{O_k\}$ constitutes the additional observables, with $\braket{O_k}$ being the estimated value from LPDO and $o_k$ the measured value from experiments.
We compare the performance of the original loss function $\Theta_0$ and the modified $\Theta^{\prime}$ in Fig.~\ref{Fig: Quafu}(b)-(c) for $100$ runs.
$\Theta_0$ cannot determine the relative phase $\varphi$ and the resulting phase and fidelity are completely random.
In contrast, the additional constraints imposed by $\Theta^{\prime}$ enable stable convergence to an accurate $\varphi$ and high fidelity.
This iteration process demonstrates the efficient and robust convergence of our algorithm on the quantum state with long-range entanglement, which is in principle more difficult to handle than previously considered random states and short-range entangled states.

For real experiments, we choose the `ScQ-P10' superconducting quantum chip with $N=10$ qubits available on the Quafu Quantum platform~\cite{Chen2022}.
The qubit configuration and noise information are shown in Fig.~\ref{Fig: Quafu}(a), where we set $L=2$ and $m=10000$, i.e., $3^L$ Pauli strings, each measured with 10,000 shots.
The residual errors using the new loss function $\Theta^{\prime}$ for different $D$ and $d_{\kappa}$ are plotted in Fig.~\ref{Fig: Quafu}(c), indicating a convergence of accuracy at $D = 4$ and $d_{\kappa}=2$, greater than those for an ideal GHZ state ($D=2$ and $d_{\kappa}=1$), implying the correlated noise effects on a noisy quantum chip in the generated quantum state.
Furthermore, the ZZ correlation function $C_{ij}^{ZZ}$ measured directly from the experiments and that estimated from the reconstructed state are compared in Fig.~\ref{Fig: Quafu}(d).
Their high consistency, especially the ability to reveal the failure of qubit 0 to correlate with other qubits, not only validates our reconstruction method but also suggests its potential to identify and mitigate errors in noisy quantum circuits~\cite{Cai2023, Guo2022}.

\section{Discussion}
In this study, we introduce a state tomography scheme that uses the LPDO parameterization of mixed states and classical data postprocessing to determine unknown tensors from only local measurements. 
Based on the ground-state witness in Eq.~\eqref{equ: witness}, our approach demonstrates enhanced efficiency, accuracy, and robustness compared to previous methods relying on MPS representations.
Meanwhile, other types of fidelity witnesses have been proposed for bosonic Gaussian states~\cite{Aolita2015} and fermionic Gaussian states~\cite{Gluza2018}, indicating the potential to generalize our method to those bosonic and fermionic states.
Specifically, some fermion Gaussian states can be mapped to the Ising model and the XX model using the Jordan-Wigner transformation~\cite{Gluza2018}, highlighting the profound connections between fidelity witnesses for different systems.
In our numerical simulations and experiments, we have considered three types of quantum states: gapped, gapless, and degenerate.
In the following, we briefly discuss the application of our method to these states.

The first class comprises the unique ground states of local, gapped Hamiltonians, such as the random state or cluster state shown in Figs.~\ref{Fig: PEPS} and~\ref{Fig: IBM}.
These states exhibit short-range correlations, with a correlation length $\xi$ that remains constant irrelevant to the system size, allowing them to be easily reconstructed from local information.
The second class includes the ground states of local but gapless Hamiltonians, such as the critical Ising and Heisenberg models studied in Figs.~\ref{Fig: Compare}-\ref{Fig: Mixed}.
These states are generally long-range correlated, with a correlation length $\xi$ comparable to the system size.
Here, `gapless' refers to an energy gap that decays according to a power law with system size, i.e., $\Delta\sim 1/N^{z}$, where $z$ is the dynamical critical exponent determined by the universality class.
Based on Eq.~\eqref{equ: witness}, we believe that such gapless states can also be reconstructed from only local marginals, though they require more experimental samplings to counteract the divergent trend induced by $\Delta$ compared to gapped states, while still maintaining a polynomial overall complexity, as indicated in Fig.~\ref{Fig: Heisenberg}(b).
The last class consists of the degenerate ground states of local, gapped Hamiltonians, such as the GHZ state considered in Fig.~\ref{Fig: Quafu}.
Although these states adhere to the entanglement area law and thus permit efficient tensor network representation, they cannot be uniquely determined by local reduced density matrices alone.
Additional global observables are necessary to complete the tomography process.
The choice of observables depends on a case-by-case analysis, necessitating some prior knowledge about the state before performing the tomography.
Fortunately, in most practical tasks aimed at benchmarking quantum hardware, the ideal state, or at least some of its fundamental properties, is known in advance, making such a modified implementation of our method feasible.

In addition, we extend TN state tomography, originally developed for 1D systems, to higher spatial dimensions.
In high-dimensional systems, besides the experiment perspective, the estimation of the tensor environment, a crucial step in the classical data postprocessing, is inherently challenging.
Approximation and truncation are generally unavoidable during contraction.
Our optimization method demonstrates efficacy in alleviating the curse of dimensionality for short-range correlated states, where the truncation error for the environment remains manageable.
Conversely, the applicability of our method to quantum states beyond short-range correlated ones warrants further investigation.
Nevertheless, the findings suggest that the recently proposed process tomography method~\cite{Torlai2023} and the error mitigation approach~\cite{Guo2022} may generalize to higher-dimensional circuits, enabling a deeper understanding of noise and its effects in such systems~\cite{Guo2023A}.
Our protocol facilitates the realization of quantum state tomography for large systems with potential quantum advantages, promoting advances in precise quantum control and complex algorithm implementation~\cite{Preskill2018, Endo2021}.

\section{Methods}
\subsection{Local measurements and reduced density matrices}
To obtain the reduced density matrix $\hat{\sigma}_{\braket{i}}$ for the sites $\{i, \cdots, i+L-1\}$ in experiments, one needs to implement an informationally complete set of measurements in this subsystem.
Specifically, we consider all possible products of Pauli operators $\hat{P}_{\braket{i}}^{\bm{m}}$ acting on these adjacent $L$ sites, where $\bm{m}=\{m_i, \cdots, m_{i+L-1}\}$ with $m_{i}\in\{I, X, Y, Z\}$.
Then $\hat{\sigma}_{\braket{i}}$ can be expanded as~\cite{Nielsen2009, Cramer2010}
\begin{equation}
    \hat{\sigma}_{\braket{i}} = \frac{1}{2^L}\sum_{\bm{m}}{\Tr{\left[\hat{\sigma}_{\braket{i}}\hat{P}_{\braket{i}}^{\bm{m}}\right]}\hat{P}_{\braket{i}}^{\bm{m}}}
\end{equation}
since $\hat{P}_{\braket{i}}^{\bm{m}}$ constitute a set of complete and orthogonal bases in the operator space.
To reconstruct each $\hat{\sigma}_{\braket{i}}$, the expectation values of all $4^L-1$ Pauli strings must be measured (as the identity operator is trivial).
In other words, a total of $(4^L-1)(N-L+1)$ observables are required to implement the tomography.

To reduce the number of different measurement settings from $(4^L-1)(N-L+1)$ to only $3^L$, we use several strategies.
First, any Pauli string containing $I$ can be reduced since its expectation value can be estimated from other Pauli strings, where $I$ is replaced by one of $\{X, Y, Z\}$.
In other words, only $3^L$ measurement bases are sufficient to reconstruct the reduced density matrix for a local subsystem of length $L$.
Second, the factor $(N-L+1)$ is effectively eliminated because a single measurement configuration with multiple shots allows the simultaneous extraction of several observable expectation values.
For example, measuring $Z$ for all qubits $Z_1\otimes Z_2\otimes\cdots\otimes Z_N$ enables the determination of all local observables in the form of $Z_{i}\otimes Z_{i+1}\otimes \cdots\otimes Z_{i+L-1}$ (product-$Z$ strings of length $L$) with sufficient samples.
In fact, all translation-invariant Pauli strings $\hat{P}_{\braket{1}}^{\bm{m}}\otimes \hat{P}_{\braket{L+1}}^{\bm{m}}\otimes \cdots \hat{P}_{\braket{N-L+1}}^{\bm{m}}$ with a period of $L$ sites, assuming that $N/L$ is an integer, are adequate to reduce to any local Pauli string of length $L$ after tracing out the corresponding environment.
Here, $\bm{m}={m_1, \cdots, m_{L}}$ and $m_{i}\in\{X, Y, Z\}$.
In essence, only $3^L$ measurement bases, each represented by a translation-invariant global Pauli string with a period of $L$, are required to compute all the necessary reduced density matrices for our tomography method.

\subsection{SVT-MPS method and ML-MPS method}
We briefly review two QST methods for 1D systems based on MPS and local measurements~\cite{Cramer2010, Baumgratz2013A}.

The SVT-MPS method~\cite{Cramer2010} is inspired by the SVT algorithm in computer science for matrix completion~\cite{Cai2008}.
The target state can be approached iteratively by solving a local Hamiltonian at each step.
Specifically, in the $n$-th iterative step we construct and find the dominant eigenstate of the following local Hamiltonian 
\begin{equation}
    \hat{Y}_{n+1} = \hat{Y}_n + \delta_n\left(\sum_i \hat{\sigma}_{\braket{i}}-E_n \sum_i \hat{\rho}_{n\braket{i}}\right).
\end{equation}
Here, $\delta_n$ is the `update rate' for each step, $\hat{\sigma}_{\braket{i}}$ are reduced density matrices for adjacent $L$ sites of the target state obtained through direct measurements, while $\hat{\rho}_{n\braket{i}}$ are reduced density matrices of the dominant eigenstate of $\hat{Y}_n$ with eigenvalue $E_n$.
The authors have shown that the number of iterative steps to achieve fixed fidelity typically scales as $O(N^2)$, where $N$ is the number of qubits.
Additionally, in each step, a variational method is needed to solve a local Hamiltonian, involving $O(N)$ sweep back and forth and $O(N)$ calculations of the environment per sweep.
Updating local tensors for each site also requires calculating the dominant eigenvector of a $D^2d_p\times D^2d_p$ matrix, with complexity at least $O(D^4)$ assuming sparsity.
Therefore, the total computational cost of SVT-MPS is $O(N^4D^4)$, which is polynomial but limited to pure states.
However, since real experiments encounter mixed states with decoherence, this reconstruction scheme has limited practicality for characterizing the noise effect in quantum devices.

The ML-MPS method directly searches for the target state that maximizes the log-likelihood function
\begin{equation}
    \log{\mathcal{L}\left(\hat{\rho}\right)} = \sum_{i, j} n_{\braket{i}}^j \log{\left(\Tr\left[\hat{\Pi}_{\braket{i}}^j\hat{\rho}\right]\right)}
\end{equation}
via a fixed-point iterative algorithm.
Here, $\hat{\Pi}_{\braket{i}}^j$ are local projectors labeled with $j$ applied at adjacent $L$ sites $\{i, \cdots, i+L-1\}$, and $n_{\braket{i}}^j$ are the corresponding measurement outcomes.
The solution $\hat{\rho}_{\textrm{ML}}$ that maximizes the above function satisfies
\begin{equation}
    \hat{\rho}_{\textrm{ML}} = \frac{1}{M}\sum_{i,j}\frac{n_{\braket{i}}^j}{\Tr{\left[\hat{\Pi}_{\braket{i}}^j\hat{\rho}_{\text{ML}}\right]}}\hat{\Pi}_{\braket{i}}^j\hat{\rho}_{\text{ML}}\equiv \mathcal{R}(\hat{\rho}_{\text{ML}})\hat{\rho}_{\text{ML}},
\end{equation}
which corresponds to the fixed-point equation
\begin{equation}
    \hat{\rho} = \mathcal{R}(\hat{\rho})\hat{\rho}\mathcal{R}(\hat{\rho}).
    \label{equ: ML-Fixed}
\end{equation}
In practice, one can replace $\mathcal{R}$ by $(\mathcal{I}+\epsilon\mathcal{R}) / (1+\epsilon)$ with $\epsilon \ll 1$.
Furthermore, under the assumption of pure state, we only need to iterate on the pure state manifold $\ket{\psi} = \mathcal{R}\ket{\psi}$.
To implement this, we construct the MPO representation of $\mathcal{R}$ in each iteration and truncate the resulting MPS $\mathcal{R}\ket{\psi}$.
Truncation can be done variationally by minimizing the error $e = (\bra{\psi}-\bra{\psi^{\prime}})(\ket{\psi}-\ket{\psi^{\prime}})$ or more efficiently by using SVD from site to site in canonical form, which requires $O(ND^{3})$ operations.

In Table~\ref{Tab: compare}, we summarize the computational complexity and application scope of the previous two methods and our Grad-LPDO method.
\begin{table}[H]
    \centering
    \caption{Computational complexity concerning system size $N$ and bond dimension $D$ and application scope for different methods.}
    \begin{tabular}{c|cc}\hline
        Method & Complexity & Application\\\hline
        SVT-MPS & $O(N^4D^4)$ & Pure states\\
        ML-MPS (MPO) & $O(\exp(N)ND^3)$ & Pure (Mixed) states \\
        Grad-LPDO & $O(\log(N)ND^3)$ & Mixed states\\\hline
    \end{tabular}
    \label{Tab: compare}
\end{table}

\subsection{Learn-LPDO method}
Here we discuss how to apply the process tomography method~\cite{Torlai2023} to the task of state tomography with LPDO.
The measurement operators are constructed as $\Pi_{\bm{\beta}}=\Pi_{\beta_1}\otimes \cdots \otimes \Pi_{\beta_N}$, where $\beta_i = 1, 2, \cdots, 6$, corresponding to $6$ projectors of one measurement randomly chosen from $\{X, Y, Z\}$ and possible outcomes.
These operators satisfy the condition $\sum_{\beta_i}{\Pi_{\beta_i}} = I$.
For any quantum state $\rho$, the probability distribution of such global random measurements is
\begin{equation}
    P_{\rho}(\bm{\beta}) = \Tr{\left[\rho \Pi_{\bm{\beta}}\right]}.
\end{equation}
Using the Kullbach-Leibler divergence between the target and reconstructed states
\begin{equation}
    D_{\rm{KL}} = \sum_{\{\bm{\beta}\}} P_{\rm{Exp}}(\bm{\beta})\log{\frac{P_{\rm{Exp}}(\bm{\beta})}{P_{\rm{LPDO}}(\bm{\beta})}},
\end{equation}
we establish the loss function to minimize as
\begin{equation}
    \Theta = -\frac{1}{M}\sum_{k=1}^{M}\log{P_{\rm{LPDO}}(\bm{\beta}_k)},
\end{equation}
where $k$ is the label of different samples with a total number of $M$.

The LPDO parameters are updated through gradient descent $A_{j} \rightarrow A_{j} - \eta\frac{\partial \Theta}{\partial A_{j}^{*}}$.
The complexity involved in calculating all gradients in one iterative step is $O(NM)$ since the loss function contains all original samples.
In our numerical simulations for Learn-LPDO, we use the Julia package PastaQ, which provides built-in functions to facilitate the sampling process and the implementation of this method.

\subsection{Fidelity between two states}
The fidelity between two normalized pure states is defined as
\begin{equation}
    f\left(\ket{\psi}, \ket{\phi}\right) = \left|\braket{\psi|\phi}\right|^2.
    \label{equ: fidelity}
\end{equation}
This is usually generalized for two mixed states as
\begin{equation}
    f_0\left(\hat{\rho}_1, \hat{\rho}_2\right) = \left(\Tr{\sqrt{\sqrt{\hat{\rho}_1}\hat{\rho}_2\sqrt{\hat{\rho}_1}}}\right)^2
\end{equation}
with normalized $\Tr{\left[\hat{\rho}_1\right]}=\Tr{\left[\hat{\rho}_2\right]}=1$.
However, this definition cannot be directly estimated for two mixed states in their LPDO form, which prevents direct benchmarking of our tomography method for mixed states in large systems.
Therefore, we adopt an alternative definition
\begin{equation}
    f\left(\hat{\rho}_1, \hat{\rho}_2\right) \equiv \Tr{\left(\hat{\rho}_1\hat{\rho}_2\right)} / \sqrt{\Tr{\left(\hat{\rho}_1^2\right)}\Tr{\left(\hat{\rho}_2^2\right)}},
\end{equation}
which is the inner product in the operator space and equals the overlap between two superoperators $\sket{\rho_1}$ and $\sket{\rho_2}$.
In particular, this alternative definition of fidelity reduces to Eq.~\eqref{equ: fidelity} for pure states.

\subsection{Noise models}
In our numerical simulations for mixed states, the noise added after each state includes four types.
The depolarizing noise is defined as
\begin{equation}
    \mathcal{E}\left(\hat{\rho}\right) = \left(1-\frac{4}{3}\varepsilon\right)\hat{\rho} + \frac{1}{3}\varepsilon\sum_{i=0}^{3}\sigma_i\hat{\rho}\sigma_i.
\end{equation}
The bit flipping noise is defined as
\begin{equation}
    \mathcal{E}\left(\hat{\rho}\right) = \left(1-\varepsilon\right)\hat{\rho} + \varepsilon\sigma_x\hat{\rho}\sigma_x.
\end{equation}
The amplitude damping noise is defined by the Kraus operator $E_0 = \ket{0}\hspace{-1mm}\bra{0} + \sqrt{1-\varepsilon}\ket{1}\hspace{-1mm}\bra{1}$ and $E_1 = \sqrt{\varepsilon}\ket{0}\hspace{-1mm}\bra{1}$ with the operator-sum representation
\begin{equation}
    \mathcal{E}\left(\hat{\rho}\right) = E_0\hat{\rho} E_0^{\dagger} + E_1\hat{\rho} E_1^{\dagger}.
\end{equation}
Phase damping noise is defined similarly, with the Kraus operator $E_0 = \ket{0}\hspace{-1mm}\bra{0} + \sqrt{1-\varepsilon}\ket{1}\hspace{-1mm}\bra{1}$ and $E_1 = \sqrt{\varepsilon}\ket{1}\hspace{-1mm}\bra{1}$.

\section{Data availability}
The datasets generated and analyzed during the current study are available from the corresponding author upon reasonable request.

\section{Code availablity}
The IBM Quantum device `ibm\_nairobi' is accessible at \href{https://quantum-computing.ibm.com}{https://quantum-computing.ibm.com/}.
The Quafu Quantum device `ScQ-P10' is accessible at \href{https://quafu.baqis.ac.cn/}{https://quafu.baqis.ac.cn/}.
The code for this study is available from the corresponding author upon reasonable request.

\bibliography{ref}

\begin{thebibliography}{87}%
\makeatletter
\providecommand \@ifxundefined [1]{%
 \@ifx{#1\undefined}
}%
\providecommand \@ifnum [1]{%
 \ifnum #1\expandafter \@firstoftwo
 \else \expandafter \@secondoftwo
 \fi
}%
\providecommand \@ifx [1]{%
 \ifx #1\expandafter \@firstoftwo
 \else \expandafter \@secondoftwo
 \fi
}%
\providecommand \natexlab [1]{#1}%
\providecommand \enquote  [1]{``#1''}%
\providecommand \bibnamefont  [1]{#1}%
\providecommand \bibfnamefont [1]{#1}%
\providecommand \citenamefont [1]{#1}%
\providecommand \href@noop [0]{\@secondoftwo}%
\providecommand \href [0]{\begingroup \@sanitize@url \@href}%
\providecommand \@href[1]{\@@startlink{#1}\@@href}%
\providecommand \@@href[1]{\endgroup#1\@@endlink}%
\providecommand \@sanitize@url [0]{\catcode `\\12\catcode `\$12\catcode
  `\&12\catcode `\#12\catcode `\^12\catcode `\_12\catcode `\%12\relax}%
\providecommand \@@startlink[1]{}%
\providecommand \@@endlink[0]{}%
\providecommand \url  [0]{\begingroup\@sanitize@url \@url }%
\providecommand \@url [1]{\endgroup\@href {#1}{\urlprefix }}%
\providecommand \urlprefix  [0]{URL }%
\providecommand \Eprint [0]{\href }%
\providecommand \doibase [0]{https://doi.org/}%
\providecommand \selectlanguage [0]{\@gobble}%
\providecommand \bibinfo  [0]{\@secondoftwo}%
\providecommand \bibfield  [0]{\@secondoftwo}%
\providecommand \translation [1]{[#1]}%
\providecommand \BibitemOpen [0]{}%
\providecommand \bibitemStop [0]{}%
\providecommand \bibitemNoStop [0]{.\EOS\space}%
\providecommand \EOS [0]{\spacefactor3000\relax}%
\providecommand \BibitemShut  [1]{\csname bibitem#1\endcsname}%
\let\auto@bib@innerbib\@empty
\bibitem [{\citenamefont {Nielsen}\ and\ \citenamefont
  {Chuang}(2009)}]{Nielsen2009}%
  \BibitemOpen
  \bibfield  {author} {\bibinfo {author} {\bibfnamefont {M.~A.}\ \bibnamefont
  {Nielsen}}\ and\ \bibinfo {author} {\bibfnamefont {I.~L.}\ \bibnamefont
  {Chuang}},\ }\href {https://doi.org/10.1017/cbo9780511976667} {\emph
  {\bibinfo {title} {Quantum Computation and Quantum Information}}}\ (\bibinfo
  {publisher} {Cambridge University Press},\ \bibinfo {year}
  {2009})\BibitemShut {NoStop}%
\bibitem [{\citenamefont {Preskill}(2018)}]{Preskill2018}%
  \BibitemOpen
  \bibfield  {author} {\bibinfo {author} {\bibfnamefont {J.}~\bibnamefont
  {Preskill}},\ }\bibfield  {title} {\bibinfo {title} {Quantum computing in the
  {NISQ} era and beyond},\ }\href {https://doi.org/10.22331/q-2018-08-06-79}
  {\bibfield  {journal} {\bibinfo  {journal} {Quantum}\ }\textbf {\bibinfo
  {volume} {2}},\ \bibinfo {pages} {79} (\bibinfo {year} {2018})}\BibitemShut
  {NoStop}%
\bibitem [{\citenamefont {Gebhart}\ \emph {et~al.}(2023)\citenamefont
  {Gebhart}, \citenamefont {Santagati}, \citenamefont {Gentile}, \citenamefont
  {Gauger}, \citenamefont {Craig}, \citenamefont {Ares}, \citenamefont
  {Banchi}, \citenamefont {Marquardt}, \citenamefont {Pezz{\`{e}}},\ and\
  \citenamefont {Bonato}}]{Gebhart2023}%
  \BibitemOpen
  \bibfield  {author} {\bibinfo {author} {\bibfnamefont {V.}~\bibnamefont
  {Gebhart}}, \bibinfo {author} {\bibfnamefont {R.}~\bibnamefont {Santagati}},
  \bibinfo {author} {\bibfnamefont {A.~A.}\ \bibnamefont {Gentile}}, \bibinfo
  {author} {\bibfnamefont {E.~M.}\ \bibnamefont {Gauger}}, \bibinfo {author}
  {\bibfnamefont {D.}~\bibnamefont {Craig}}, \bibinfo {author} {\bibfnamefont
  {N.}~\bibnamefont {Ares}}, \bibinfo {author} {\bibfnamefont {L.}~\bibnamefont
  {Banchi}}, \bibinfo {author} {\bibfnamefont {F.}~\bibnamefont {Marquardt}},
  \bibinfo {author} {\bibfnamefont {L.}~\bibnamefont {Pezz{\`{e}}}},\ and\
  \bibinfo {author} {\bibfnamefont {C.}~\bibnamefont {Bonato}},\ }\bibfield
  {title} {\bibinfo {title} {Learning quantum systems},\ }\href
  {https://doi.org/10.1038/s42254-022-00552-1} {\bibfield  {journal} {\bibinfo
  {journal} {Nat. Rev. Phys.}\ }\textbf {\bibinfo {volume} {5}},\ \bibinfo
  {pages} {141} (\bibinfo {year} {2023})}\BibitemShut {NoStop}%
\bibitem [{\citenamefont {Arute}\ \emph {et~al.}(2019)\citenamefont {Arute},
  \citenamefont {Arya}, \citenamefont {Babbush}, \citenamefont {Bacon} \emph
  {et~al.}}]{Arute2019}%
  \BibitemOpen
  \bibfield  {author} {\bibinfo {author} {\bibfnamefont {F.}~\bibnamefont
  {Arute}}, \bibinfo {author} {\bibfnamefont {K.}~\bibnamefont {Arya}},
  \bibinfo {author} {\bibfnamefont {R.}~\bibnamefont {Babbush}}, \bibinfo
  {author} {\bibfnamefont {D.}~\bibnamefont {Bacon}}, \emph {et~al.},\
  }\bibfield  {title} {\bibinfo {title} {Quantum supremacy using a programmable
  superconducting processor},\ }\href
  {https://doi.org/10.1038/s41586-019-1666-5} {\bibfield  {journal} {\bibinfo
  {journal} {Nature}\ }\textbf {\bibinfo {volume} {574}},\ \bibinfo {pages}
  {505} (\bibinfo {year} {2019})}\BibitemShut {NoStop}%
\bibitem [{\citenamefont {Kim}\ \emph {et~al.}(2023)\citenamefont {Kim},
  \citenamefont {Eddins}, \citenamefont {Anand}, \citenamefont {Wei},
  \citenamefont {van~den Berg}, \citenamefont {Rosenblatt}, \citenamefont
  {Nayfeh}, \citenamefont {Wu}, \citenamefont {Zaletel}, \citenamefont
  {Temme},\ and\ \citenamefont {Kandala}}]{Kim2023}%
  \BibitemOpen
  \bibfield  {author} {\bibinfo {author} {\bibfnamefont {Y.}~\bibnamefont
  {Kim}}, \bibinfo {author} {\bibfnamefont {A.}~\bibnamefont {Eddins}},
  \bibinfo {author} {\bibfnamefont {S.}~\bibnamefont {Anand}}, \bibinfo
  {author} {\bibfnamefont {K.~X.}\ \bibnamefont {Wei}}, \bibinfo {author}
  {\bibfnamefont {E.}~\bibnamefont {van~den Berg}}, \bibinfo {author}
  {\bibfnamefont {S.}~\bibnamefont {Rosenblatt}}, \bibinfo {author}
  {\bibfnamefont {H.}~\bibnamefont {Nayfeh}}, \bibinfo {author} {\bibfnamefont
  {Y.}~\bibnamefont {Wu}}, \bibinfo {author} {\bibfnamefont {M.}~\bibnamefont
  {Zaletel}}, \bibinfo {author} {\bibfnamefont {K.}~\bibnamefont {Temme}},\
  and\ \bibinfo {author} {\bibfnamefont {A.}~\bibnamefont {Kandala}},\
  }\bibfield  {title} {\bibinfo {title} {Evidence for the utility of quantum
  computing before fault tolerance},\ }\href
  {https://doi.org/10.1038/s41586-023-06096-3} {\bibfield  {journal} {\bibinfo
  {journal} {Nature}\ }\textbf {\bibinfo {volume} {618}},\ \bibinfo {pages}
  {500} (\bibinfo {year} {2023})}\BibitemShut {NoStop}%
\bibitem [{\citenamefont {Verstraete}\ and\ \citenamefont
  {Cirac}(2006)}]{Verstraete2006A}%
  \BibitemOpen
  \bibfield  {author} {\bibinfo {author} {\bibfnamefont {F.}~\bibnamefont
  {Verstraete}}\ and\ \bibinfo {author} {\bibfnamefont {J.~I.}\ \bibnamefont
  {Cirac}},\ }\bibfield  {title} {\bibinfo {title} {Matrix product states
  represent ground states faithfully},\ }\href
  {https://doi.org/10.1103/PhysRevB.73.094423} {\bibfield  {journal} {\bibinfo
  {journal} {Phys. Rev. B}\ }\textbf {\bibinfo {volume} {73}},\ \bibinfo
  {pages} {094423} (\bibinfo {year} {2006})}\BibitemShut {NoStop}%
\bibitem [{\citenamefont {P\'erez-Garc\'{\i}a}\ \emph
  {et~al.}(2007)\citenamefont {P\'erez-Garc\'{\i}a}, \citenamefont
  {Verstraete}, \citenamefont {Wolf},\ and\ \citenamefont {Cirac}}]{Perez2007}%
  \BibitemOpen
  \bibfield  {author} {\bibinfo {author} {\bibfnamefont {D.}~\bibnamefont
  {P\'erez-Garc\'{\i}a}}, \bibinfo {author} {\bibfnamefont {F.}~\bibnamefont
  {Verstraete}}, \bibinfo {author} {\bibfnamefont {M.~M.}\ \bibnamefont
  {Wolf}},\ and\ \bibinfo {author} {\bibfnamefont {J.~I.}\ \bibnamefont
  {Cirac}},\ }\bibfield  {title} {\bibinfo {title} {Matrix product state
  representations},\ }\href {https://doi.org/10.26421/QIC7.5-6-1} {\bibfield
  {journal} {\bibinfo  {journal} {Quantum Info. Comput.}\ }\textbf {\bibinfo
  {volume} {7}},\ \bibinfo {pages} {401–430} (\bibinfo {year}
  {2007})}\BibitemShut {NoStop}%
\bibitem [{\citenamefont {Verstraete}\ \emph {et~al.}(2008)\citenamefont
  {Verstraete}, \citenamefont {Murg},\ and\ \citenamefont
  {Cirac}}]{Verstraete2008}%
  \BibitemOpen
  \bibfield  {author} {\bibinfo {author} {\bibfnamefont {F.}~\bibnamefont
  {Verstraete}}, \bibinfo {author} {\bibfnamefont {V.}~\bibnamefont {Murg}},\
  and\ \bibinfo {author} {\bibfnamefont {J.}~\bibnamefont {Cirac}},\ }\bibfield
   {title} {\bibinfo {title} {Matrix product states, projected entangled pair
  states, and variational renormalization group methods for quantum spin
  systems},\ }\href {https://doi.org/10.1080/14789940801912366} {\bibfield
  {journal} {\bibinfo  {journal} {Adv. Phys.}\ }\textbf {\bibinfo {volume}
  {57}},\ \bibinfo {pages} {143} (\bibinfo {year} {2008})}\BibitemShut
  {NoStop}%
\bibitem [{\citenamefont {Schollwöck}(2011)}]{Schollwoeck2011}%
  \BibitemOpen
  \bibfield  {author} {\bibinfo {author} {\bibfnamefont {U.}~\bibnamefont
  {Schollwöck}},\ }\bibfield  {title} {\bibinfo {title} {The density-matrix
  renormalization group in the age of matrix product states},\ }\href
  {https://doi.org/10.1016/j.aop.2010.09.012} {\bibfield  {journal} {\bibinfo
  {journal} {Ann. Phys.}\ }\textbf {\bibinfo {volume} {326}},\ \bibinfo {pages}
  {96} (\bibinfo {year} {2011})}\BibitemShut {NoStop}%
\bibitem [{\citenamefont {Orús}(2014)}]{Orus2014}%
  \BibitemOpen
  \bibfield  {author} {\bibinfo {author} {\bibfnamefont {R.}~\bibnamefont
  {Orús}},\ }\bibfield  {title} {\bibinfo {title} {A practical introduction to
  tensor networks: Matrix product states and projected entangled pair states},\
  }\href {https://doi.org/10.1016/j.aop.2014.06.013} {\bibfield  {journal}
  {\bibinfo  {journal} {Ann. Phys.}\ }\textbf {\bibinfo {volume} {349}},\
  \bibinfo {pages} {117} (\bibinfo {year} {2014})}\BibitemShut {NoStop}%
\bibitem [{\citenamefont {Cirac}\ \emph {et~al.}(2021)\citenamefont {Cirac},
  \citenamefont {P\'erez-Garc\'{\i}a}, \citenamefont {Schuch},\ and\
  \citenamefont {Verstraete}}]{Cirac2021}%
  \BibitemOpen
  \bibfield  {author} {\bibinfo {author} {\bibfnamefont {J.~I.}\ \bibnamefont
  {Cirac}}, \bibinfo {author} {\bibfnamefont {D.}~\bibnamefont
  {P\'erez-Garc\'{\i}a}}, \bibinfo {author} {\bibfnamefont {N.}~\bibnamefont
  {Schuch}},\ and\ \bibinfo {author} {\bibfnamefont {F.}~\bibnamefont
  {Verstraete}},\ }\bibfield  {title} {\bibinfo {title} {Matrix product states
  and projected entangled pair states: Concepts, symmetries, theorems},\ }\href
  {https://doi.org/10.1103/RevModPhys.93.045003} {\bibfield  {journal}
  {\bibinfo  {journal} {Rev. Mod. Phys.}\ }\textbf {\bibinfo {volume} {93}},\
  \bibinfo {pages} {045003} (\bibinfo {year} {2021})}\BibitemShut {NoStop}%
\bibitem [{\citenamefont {Verstraete}\ \emph {et~al.}(2023)\citenamefont
  {Verstraete}, \citenamefont {Nishino}, \citenamefont {Schollwöck},
  \citenamefont {Ba{\~{n}}uls}, \citenamefont {Chan},\ and\ \citenamefont
  {Stoudenmire}}]{Verstraete2023}%
  \BibitemOpen
  \bibfield  {author} {\bibinfo {author} {\bibfnamefont {F.}~\bibnamefont
  {Verstraete}}, \bibinfo {author} {\bibfnamefont {T.}~\bibnamefont {Nishino}},
  \bibinfo {author} {\bibfnamefont {U.}~\bibnamefont {Schollwöck}}, \bibinfo
  {author} {\bibfnamefont {M.~C.}\ \bibnamefont {Ba{\~{n}}uls}}, \bibinfo
  {author} {\bibfnamefont {G.~K.}\ \bibnamefont {Chan}},\ and\ \bibinfo
  {author} {\bibfnamefont {M.~E.}\ \bibnamefont {Stoudenmire}},\ }\bibfield
  {title} {\bibinfo {title} {Density matrix renormalization group, 30 years
  on},\ }\href {https://doi.org/10.1038/s42254-023-00572-5} {\bibfield
  {journal} {\bibinfo  {journal} {Nat. Rev. Phys.}\ }\textbf {\bibinfo {volume}
  {5}},\ \bibinfo {pages} {273} (\bibinfo {year} {2023})}\BibitemShut {NoStop}%
\bibitem [{\citenamefont {Cramer}\ \emph {et~al.}(2010)\citenamefont {Cramer},
  \citenamefont {Plenio}, \citenamefont {Flammia}, \citenamefont {Somma},
  \citenamefont {Gross}, \citenamefont {Bartlett}, \citenamefont
  {Landon-Cardinal}, \citenamefont {Poulin},\ and\ \citenamefont
  {Liu}}]{Cramer2010}%
  \BibitemOpen
  \bibfield  {author} {\bibinfo {author} {\bibfnamefont {M.}~\bibnamefont
  {Cramer}}, \bibinfo {author} {\bibfnamefont {M.~B.}\ \bibnamefont {Plenio}},
  \bibinfo {author} {\bibfnamefont {S.~T.}\ \bibnamefont {Flammia}}, \bibinfo
  {author} {\bibfnamefont {R.}~\bibnamefont {Somma}}, \bibinfo {author}
  {\bibfnamefont {D.}~\bibnamefont {Gross}}, \bibinfo {author} {\bibfnamefont
  {S.~D.}\ \bibnamefont {Bartlett}}, \bibinfo {author} {\bibfnamefont
  {O.}~\bibnamefont {Landon-Cardinal}}, \bibinfo {author} {\bibfnamefont
  {D.}~\bibnamefont {Poulin}},\ and\ \bibinfo {author} {\bibfnamefont {Y.-K.}\
  \bibnamefont {Liu}},\ }\bibfield  {title} {\bibinfo {title} {Efficient
  quantum state tomography},\ }\href {https://doi.org/10.1038/ncomms1147}
  {\bibfield  {journal} {\bibinfo  {journal} {Nat. Commun.}\ }\textbf {\bibinfo
  {volume} {1}},\ \bibinfo {pages} {149} (\bibinfo {year} {2010})}\BibitemShut
  {NoStop}%
\bibitem [{\citenamefont {Baumgratz}\ \emph
  {et~al.}(2013{\natexlab{a}})\citenamefont {Baumgratz}, \citenamefont
  {Nüßeler}, \citenamefont {Cramer},\ and\ \citenamefont
  {Plenio}}]{Baumgratz2013A}%
  \BibitemOpen
  \bibfield  {author} {\bibinfo {author} {\bibfnamefont {T.}~\bibnamefont
  {Baumgratz}}, \bibinfo {author} {\bibfnamefont {A.}~\bibnamefont
  {Nüßeler}}, \bibinfo {author} {\bibfnamefont {M.}~\bibnamefont {Cramer}},\
  and\ \bibinfo {author} {\bibfnamefont {M.~B.}\ \bibnamefont {Plenio}},\
  }\bibfield  {title} {\bibinfo {title} {A scalable maximum likelihood method
  for quantum state tomography},\ }\href
  {https://doi.org/10.1088/1367-2630/15/12/125004} {\bibfield  {journal}
  {\bibinfo  {journal} {New J. Phys.}\ }\textbf {\bibinfo {volume} {15}},\
  \bibinfo {pages} {125004} (\bibinfo {year} {2013}{\natexlab{a}})}\BibitemShut
  {NoStop}%
\bibitem [{\citenamefont {Lanyon}\ \emph {et~al.}(2017)\citenamefont {Lanyon},
  \citenamefont {Maier}, \citenamefont {Holzäpfel}, \citenamefont {Baumgratz},
  \citenamefont {Hempel}, \citenamefont {Jurcevic}, \citenamefont {Dhand},
  \citenamefont {Buyskikh}, \citenamefont {Daley}, \citenamefont {Cramer},
  \citenamefont {Plenio}, \citenamefont {Blatt},\ and\ \citenamefont
  {Roos}}]{Lanyon2017}%
  \BibitemOpen
  \bibfield  {author} {\bibinfo {author} {\bibfnamefont {B.~P.}\ \bibnamefont
  {Lanyon}}, \bibinfo {author} {\bibfnamefont {C.}~\bibnamefont {Maier}},
  \bibinfo {author} {\bibfnamefont {M.}~\bibnamefont {Holzäpfel}}, \bibinfo
  {author} {\bibfnamefont {T.}~\bibnamefont {Baumgratz}}, \bibinfo {author}
  {\bibfnamefont {C.}~\bibnamefont {Hempel}}, \bibinfo {author} {\bibfnamefont
  {P.}~\bibnamefont {Jurcevic}}, \bibinfo {author} {\bibfnamefont
  {I.}~\bibnamefont {Dhand}}, \bibinfo {author} {\bibfnamefont {A.~S.}\
  \bibnamefont {Buyskikh}}, \bibinfo {author} {\bibfnamefont {A.~J.}\
  \bibnamefont {Daley}}, \bibinfo {author} {\bibfnamefont {M.}~\bibnamefont
  {Cramer}}, \bibinfo {author} {\bibfnamefont {M.~B.}\ \bibnamefont {Plenio}},
  \bibinfo {author} {\bibfnamefont {R.}~\bibnamefont {Blatt}},\ and\ \bibinfo
  {author} {\bibfnamefont {C.~F.}\ \bibnamefont {Roos}},\ }\bibfield  {title}
  {\bibinfo {title} {Efficient tomography of a quantum many-body~system},\
  }\href {https://doi.org/10.1038/nphys4244} {\bibfield  {journal} {\bibinfo
  {journal} {Nat. Phys.}\ }\textbf {\bibinfo {volume} {13}},\ \bibinfo {pages}
  {1158} (\bibinfo {year} {2017})}\BibitemShut {NoStop}%
\bibitem [{\citenamefont {Pirvu}\ \emph {et~al.}(2010)\citenamefont {Pirvu},
  \citenamefont {Murg}, \citenamefont {Cirac},\ and\ \citenamefont
  {Verstraete}}]{Pirvu2010}%
  \BibitemOpen
  \bibfield  {author} {\bibinfo {author} {\bibfnamefont {B.}~\bibnamefont
  {Pirvu}}, \bibinfo {author} {\bibfnamefont {V.}~\bibnamefont {Murg}},
  \bibinfo {author} {\bibfnamefont {J.~I.}\ \bibnamefont {Cirac}},\ and\
  \bibinfo {author} {\bibfnamefont {F.}~\bibnamefont {Verstraete}},\ }\bibfield
   {title} {\bibinfo {title} {Matrix product operator representations},\ }\href
  {https://doi.org/10.1088/1367-2630/12/2/025012} {\bibfield  {journal}
  {\bibinfo  {journal} {New J. Phys.}\ }\textbf {\bibinfo {volume} {12}},\
  \bibinfo {pages} {025012} (\bibinfo {year} {2010})}\BibitemShut {NoStop}%
\bibitem [{\citenamefont {Guth~Jarkovsk\'y}\ \emph {et~al.}(2020)\citenamefont
  {Guth~Jarkovsk\'y}, \citenamefont {Moln\'ar}, \citenamefont {Schuch},\ and\
  \citenamefont {Cirac}}]{Jarkovsky2020}%
  \BibitemOpen
  \bibfield  {author} {\bibinfo {author} {\bibfnamefont {J.}~\bibnamefont
  {Guth~Jarkovsk\'y}}, \bibinfo {author} {\bibfnamefont {A.}~\bibnamefont
  {Moln\'ar}}, \bibinfo {author} {\bibfnamefont {N.}~\bibnamefont {Schuch}},\
  and\ \bibinfo {author} {\bibfnamefont {J.~I.}\ \bibnamefont {Cirac}},\
  }\bibfield  {title} {\bibinfo {title} {Efficient description of many-body
  systems with matrix product density operators},\ }\href
  {https://doi.org/10.1103/PRXQuantum.1.010304} {\bibfield  {journal} {\bibinfo
   {journal} {PRX Quantum}\ }\textbf {\bibinfo {volume} {1}},\ \bibinfo {pages}
  {010304} (\bibinfo {year} {2020})}\BibitemShut {NoStop}%
\bibitem [{\citenamefont {Baumgratz}\ \emph
  {et~al.}(2013{\natexlab{b}})\citenamefont {Baumgratz}, \citenamefont {Gross},
  \citenamefont {Cramer},\ and\ \citenamefont {Plenio}}]{Baumgratz2013B}%
  \BibitemOpen
  \bibfield  {author} {\bibinfo {author} {\bibfnamefont {T.}~\bibnamefont
  {Baumgratz}}, \bibinfo {author} {\bibfnamefont {D.}~\bibnamefont {Gross}},
  \bibinfo {author} {\bibfnamefont {M.}~\bibnamefont {Cramer}},\ and\ \bibinfo
  {author} {\bibfnamefont {M.~B.}\ \bibnamefont {Plenio}},\ }\bibfield  {title}
  {\bibinfo {title} {Scalable reconstruction of density matrices},\ }\href
  {https://doi.org/10.1103/PhysRevLett.111.020401} {\bibfield  {journal}
  {\bibinfo  {journal} {Phys. Rev. Lett.}\ }\textbf {\bibinfo {volume} {111}},\
  \bibinfo {pages} {020401} (\bibinfo {year} {2013}{\natexlab{b}})}\BibitemShut
  {NoStop}%
\bibitem [{\citenamefont {Han}\ \emph {et~al.}(2018)\citenamefont {Han},
  \citenamefont {Wang}, \citenamefont {Fan}, \citenamefont {Wang},\ and\
  \citenamefont {Zhang}}]{Han2018}%
  \BibitemOpen
  \bibfield  {author} {\bibinfo {author} {\bibfnamefont {Z.-Y.}\ \bibnamefont
  {Han}}, \bibinfo {author} {\bibfnamefont {J.}~\bibnamefont {Wang}}, \bibinfo
  {author} {\bibfnamefont {H.}~\bibnamefont {Fan}}, \bibinfo {author}
  {\bibfnamefont {L.}~\bibnamefont {Wang}},\ and\ \bibinfo {author}
  {\bibfnamefont {P.}~\bibnamefont {Zhang}},\ }\bibfield  {title} {\bibinfo
  {title} {Unsupervised generative modeling using matrix product states},\
  }\href {https://doi.org/10.1103/PhysRevX.8.031012} {\bibfield  {journal}
  {\bibinfo  {journal} {Phys. Rev. X}\ }\textbf {\bibinfo {volume} {8}},\
  \bibinfo {pages} {031012} (\bibinfo {year} {2018})}\BibitemShut {NoStop}%
\bibitem [{\citenamefont {Wang}\ \emph {et~al.}(2020)\citenamefont {Wang},
  \citenamefont {Han}, \citenamefont {Wang}, \citenamefont {Li}, \citenamefont
  {Mu}, \citenamefont {Fan},\ and\ \citenamefont {Wang}}]{Wang2020}%
  \BibitemOpen
  \bibfield  {author} {\bibinfo {author} {\bibfnamefont {J.}~\bibnamefont
  {Wang}}, \bibinfo {author} {\bibfnamefont {Z.-Y.}\ \bibnamefont {Han}},
  \bibinfo {author} {\bibfnamefont {S.-B.}\ \bibnamefont {Wang}}, \bibinfo
  {author} {\bibfnamefont {Z.}~\bibnamefont {Li}}, \bibinfo {author}
  {\bibfnamefont {L.-Z.}\ \bibnamefont {Mu}}, \bibinfo {author} {\bibfnamefont
  {H.}~\bibnamefont {Fan}},\ and\ \bibinfo {author} {\bibfnamefont
  {L.}~\bibnamefont {Wang}},\ }\bibfield  {title} {\bibinfo {title} {Scalable
  quantum tomography with fidelity estimation},\ }\href
  {https://doi.org/10.1103/PhysRevA.101.032321} {\bibfield  {journal} {\bibinfo
   {journal} {Phys. Rev. A}\ }\textbf {\bibinfo {volume} {101}},\ \bibinfo
  {pages} {032321} (\bibinfo {year} {2020})}\BibitemShut {NoStop}%
\bibitem [{\citenamefont {Torlai}\ \emph {et~al.}(2023)\citenamefont {Torlai},
  \citenamefont {Wood}, \citenamefont {Acharya}, \citenamefont {Carleo},
  \citenamefont {Carrasquilla},\ and\ \citenamefont {Aolita}}]{Torlai2023}%
  \BibitemOpen
  \bibfield  {author} {\bibinfo {author} {\bibfnamefont {G.}~\bibnamefont
  {Torlai}}, \bibinfo {author} {\bibfnamefont {C.~J.}\ \bibnamefont {Wood}},
  \bibinfo {author} {\bibfnamefont {A.}~\bibnamefont {Acharya}}, \bibinfo
  {author} {\bibfnamefont {G.}~\bibnamefont {Carleo}}, \bibinfo {author}
  {\bibfnamefont {J.}~\bibnamefont {Carrasquilla}},\ and\ \bibinfo {author}
  {\bibfnamefont {L.}~\bibnamefont {Aolita}},\ }\bibfield  {title} {\bibinfo
  {title} {Quantum process tomography with unsupervised learning and tensor
  networks},\ }\href {https://doi.org/10.1038/s41467-023-38332-9} {\bibfield
  {journal} {\bibinfo  {journal} {Nat. Commun.}\ }\textbf {\bibinfo {volume}
  {14}},\ \bibinfo {pages} {2858} (\bibinfo {year} {2023})}\BibitemShut
  {NoStop}%
\bibitem [{\citenamefont {Jamiołkowski}(1972)}]{Jamiolkowski1972}%
  \BibitemOpen
  \bibfield  {author} {\bibinfo {author} {\bibfnamefont {A.}~\bibnamefont
  {Jamiołkowski}},\ }\bibfield  {title} {\bibinfo {title} {Linear
  transformations which preserve trace and positive semidefiniteness of
  operators},\ }\href
  {https://doi.org/https://doi.org/10.1016/0034-4877(72)90011-0} {\bibfield
  {journal} {\bibinfo  {journal} {Rep. Math. Phys.}\ }\textbf {\bibinfo
  {volume} {3}},\ \bibinfo {pages} {275} (\bibinfo {year} {1972})}\BibitemShut
  {NoStop}%
\bibitem [{\citenamefont {Choi}(1975)}]{Choi1975}%
  \BibitemOpen
  \bibfield  {author} {\bibinfo {author} {\bibfnamefont {M.-D.}\ \bibnamefont
  {Choi}},\ }\bibfield  {title} {\bibinfo {title} {Completely positive linear
  maps on complex matrices},\ }\href
  {https://doi.org/https://doi.org/10.1016/0024-3795(75)90075-0} {\bibfield
  {journal} {\bibinfo  {journal} {Linear Alg. Appl.}\ }\textbf {\bibinfo
  {volume} {10}},\ \bibinfo {pages} {285} (\bibinfo {year} {1975})}\BibitemShut
  {NoStop}%
\bibitem [{\citenamefont {Li}\ and\ \citenamefont {Benjamin}(2017)}]{Li2017}%
  \BibitemOpen
  \bibfield  {author} {\bibinfo {author} {\bibfnamefont {Y.}~\bibnamefont
  {Li}}\ and\ \bibinfo {author} {\bibfnamefont {S.~C.}\ \bibnamefont
  {Benjamin}},\ }\bibfield  {title} {\bibinfo {title} {Efficient variational
  quantum simulator incorporating active error minimization},\ }\href
  {https://doi.org/10.1103/physrevx.7.021050} {\bibfield  {journal} {\bibinfo
  {journal} {Phys. Rev. X}\ }\textbf {\bibinfo {volume} {7}},\ \bibinfo {pages}
  {021050} (\bibinfo {year} {2017})}\BibitemShut {NoStop}%
\bibitem [{\citenamefont {Temme}\ \emph {et~al.}(2017)\citenamefont {Temme},
  \citenamefont {Bravyi},\ and\ \citenamefont {Gambetta}}]{Temme2017}%
  \BibitemOpen
  \bibfield  {author} {\bibinfo {author} {\bibfnamefont {K.}~\bibnamefont
  {Temme}}, \bibinfo {author} {\bibfnamefont {S.}~\bibnamefont {Bravyi}},\ and\
  \bibinfo {author} {\bibfnamefont {J.~M.}\ \bibnamefont {Gambetta}},\
  }\bibfield  {title} {\bibinfo {title} {Error mitigation for short-depth
  quantum circuits},\ }\href {https://doi.org/10.1103/physrevlett.119.180509}
  {\bibfield  {journal} {\bibinfo  {journal} {Phys. Rev. Lett.}\ }\textbf
  {\bibinfo {volume} {119}},\ \bibinfo {pages} {180509} (\bibinfo {year}
  {2017})}\BibitemShut {NoStop}%
\bibitem [{\citenamefont {Endo}\ \emph {et~al.}(2018)\citenamefont {Endo},
  \citenamefont {Benjamin},\ and\ \citenamefont {Li}}]{Endo2018}%
  \BibitemOpen
  \bibfield  {author} {\bibinfo {author} {\bibfnamefont {S.}~\bibnamefont
  {Endo}}, \bibinfo {author} {\bibfnamefont {S.~C.}\ \bibnamefont {Benjamin}},\
  and\ \bibinfo {author} {\bibfnamefont {Y.}~\bibnamefont {Li}},\ }\bibfield
  {title} {\bibinfo {title} {Practical quantum error mitigation for near-future
  applications},\ }\href {https://doi.org/10.1103/PhysRevX.8.031027} {\bibfield
   {journal} {\bibinfo  {journal} {Phys. Rev. X}\ }\textbf {\bibinfo {volume}
  {8}},\ \bibinfo {pages} {031027} (\bibinfo {year} {2018})}\BibitemShut
  {NoStop}%
\bibitem [{\citenamefont {McArdle}\ \emph {et~al.}(2019)\citenamefont
  {McArdle}, \citenamefont {Yuan},\ and\ \citenamefont
  {Benjamin}}]{McArdle2019}%
  \BibitemOpen
  \bibfield  {author} {\bibinfo {author} {\bibfnamefont {S.}~\bibnamefont
  {McArdle}}, \bibinfo {author} {\bibfnamefont {X.}~\bibnamefont {Yuan}},\ and\
  \bibinfo {author} {\bibfnamefont {S.}~\bibnamefont {Benjamin}},\ }\bibfield
  {title} {\bibinfo {title} {Error-mitigated digital quantum simulation},\
  }\href {https://doi.org/10.1103/physrevlett.122.180501} {\bibfield  {journal}
  {\bibinfo  {journal} {Phys. Rev. Lett.}\ }\textbf {\bibinfo {volume} {122}},\
  \bibinfo {pages} {180501} (\bibinfo {year} {2019})}\BibitemShut {NoStop}%
\bibitem [{\citenamefont {Cai}(2021)}]{Cai2021}%
  \BibitemOpen
  \bibfield  {author} {\bibinfo {author} {\bibfnamefont {Z.}~\bibnamefont
  {Cai}},\ }\bibfield  {title} {\bibinfo {title} {Multi-exponential error
  extrapolation and combining error mitigation techniques for {NISQ}
  applications},\ }\href {https://doi.org/10.1038/s41534-021-00404-3}
  {\bibfield  {journal} {\bibinfo  {journal} {npj Quantum Inform.}\ }\textbf
  {\bibinfo {volume} {7}},\ \bibinfo {pages} {80} (\bibinfo {year}
  {2021})}\BibitemShut {NoStop}%
\bibitem [{\citenamefont {Guo}\ and\ \citenamefont {Yang}(2022)}]{Guo2022}%
  \BibitemOpen
  \bibfield  {author} {\bibinfo {author} {\bibfnamefont {Y.}~\bibnamefont
  {Guo}}\ and\ \bibinfo {author} {\bibfnamefont {S.}~\bibnamefont {Yang}},\
  }\bibfield  {title} {\bibinfo {title} {Quantum error mitigation via matrix
  product operators},\ }\href {https://doi.org/10.1103/PRXQuantum.3.040313}
  {\bibfield  {journal} {\bibinfo  {journal} {PRX Quantum}\ }\textbf {\bibinfo
  {volume} {3}},\ \bibinfo {pages} {040313} (\bibinfo {year}
  {2022})}\BibitemShut {NoStop}%
\bibitem [{\citenamefont {van~den Berg}\ \emph {et~al.}(2023)\citenamefont
  {van~den Berg}, \citenamefont {Minev}, \citenamefont {Kandala},\ and\
  \citenamefont {Temme}}]{Berg2023}%
  \BibitemOpen
  \bibfield  {author} {\bibinfo {author} {\bibfnamefont {E.}~\bibnamefont
  {van~den Berg}}, \bibinfo {author} {\bibfnamefont {Z.~K.}\ \bibnamefont
  {Minev}}, \bibinfo {author} {\bibfnamefont {A.}~\bibnamefont {Kandala}},\
  and\ \bibinfo {author} {\bibfnamefont {K.}~\bibnamefont {Temme}},\ }\bibfield
   {title} {\bibinfo {title} {Probabilistic error cancellation with sparse
  pauli{\textendash}lindblad models on noisy quantum processors},\ }\href
  {https://doi.org/10.1038/s41567-023-02042-2} {\bibfield  {journal} {\bibinfo
  {journal} {Nat. Phys.}\ }\textbf {\bibinfo {volume} {19}},\ \bibinfo {pages}
  {1116} (\bibinfo {year} {2023})}\BibitemShut {NoStop}%
\bibitem [{\citenamefont {Cai}\ \emph {et~al.}(2023)\citenamefont {Cai},
  \citenamefont {Babbush}, \citenamefont {Benjamin}, \citenamefont {Endo},
  \citenamefont {Huggins}, \citenamefont {Li}, \citenamefont {McClean},\ and\
  \citenamefont {O'Brien}}]{Cai2023}%
  \BibitemOpen
  \bibfield  {author} {\bibinfo {author} {\bibfnamefont {Z.}~\bibnamefont
  {Cai}}, \bibinfo {author} {\bibfnamefont {R.}~\bibnamefont {Babbush}},
  \bibinfo {author} {\bibfnamefont {S.~C.}\ \bibnamefont {Benjamin}}, \bibinfo
  {author} {\bibfnamefont {S.}~\bibnamefont {Endo}}, \bibinfo {author}
  {\bibfnamefont {W.~J.}\ \bibnamefont {Huggins}}, \bibinfo {author}
  {\bibfnamefont {Y.}~\bibnamefont {Li}}, \bibinfo {author} {\bibfnamefont
  {J.~R.}\ \bibnamefont {McClean}},\ and\ \bibinfo {author} {\bibfnamefont
  {T.~E.}\ \bibnamefont {O'Brien}},\ }\href@noop {} {\bibinfo {title} {Quantum
  error mitigation}} (\bibinfo {year} {2023}),\ \Eprint
  {https://arxiv.org/abs/2210.00921} {arXiv:2210.00921} \BibitemShut {NoStop}%
\bibitem [{\citenamefont {van~den Berg}\ \emph {et~al.}(2022)\citenamefont
  {van~den Berg}, \citenamefont {Minev},\ and\ \citenamefont
  {Temme}}]{Berg2022}%
  \BibitemOpen
  \bibfield  {author} {\bibinfo {author} {\bibfnamefont {E.}~\bibnamefont
  {van~den Berg}}, \bibinfo {author} {\bibfnamefont {Z.~K.}\ \bibnamefont
  {Minev}},\ and\ \bibinfo {author} {\bibfnamefont {K.}~\bibnamefont {Temme}},\
  }\bibfield  {title} {\bibinfo {title} {Model-free readout-error mitigation
  for quantum expectation values},\ }\href
  {https://doi.org/10.1103/PhysRevA.105.032620} {\bibfield  {journal} {\bibinfo
   {journal} {Phys. Rev. A}\ }\textbf {\bibinfo {volume} {105}},\ \bibinfo
  {pages} {032620} (\bibinfo {year} {2022})}\BibitemShut {NoStop}%
\bibitem [{\citenamefont {Yang}\ \emph {et~al.}(2022)\citenamefont {Yang},
  \citenamefont {Raymond},\ and\ \citenamefont {Uno}}]{Yang2022}%
  \BibitemOpen
  \bibfield  {author} {\bibinfo {author} {\bibfnamefont {B.}~\bibnamefont
  {Yang}}, \bibinfo {author} {\bibfnamefont {R.}~\bibnamefont {Raymond}},\ and\
  \bibinfo {author} {\bibfnamefont {S.}~\bibnamefont {Uno}},\ }\bibfield
  {title} {\bibinfo {title} {Efficient quantum readout-error mitigation for
  sparse measurement outcomes of near-term quantum devices},\ }\href
  {https://doi.org/10.1103/PhysRevA.106.012423} {\bibfield  {journal} {\bibinfo
   {journal} {Phys. Rev. A}\ }\textbf {\bibinfo {volume} {106}},\ \bibinfo
  {pages} {012423} (\bibinfo {year} {2022})}\BibitemShut {NoStop}%
\bibitem [{\citenamefont {Torlai}\ \emph {et~al.}(2018)\citenamefont {Torlai},
  \citenamefont {Mazzola}, \citenamefont {Carrasquilla}, \citenamefont
  {Troyer}, \citenamefont {Melko},\ and\ \citenamefont {Carleo}}]{Torlai2018}%
  \BibitemOpen
  \bibfield  {author} {\bibinfo {author} {\bibfnamefont {G.}~\bibnamefont
  {Torlai}}, \bibinfo {author} {\bibfnamefont {G.}~\bibnamefont {Mazzola}},
  \bibinfo {author} {\bibfnamefont {J.}~\bibnamefont {Carrasquilla}}, \bibinfo
  {author} {\bibfnamefont {M.}~\bibnamefont {Troyer}}, \bibinfo {author}
  {\bibfnamefont {R.}~\bibnamefont {Melko}},\ and\ \bibinfo {author}
  {\bibfnamefont {G.}~\bibnamefont {Carleo}},\ }\bibfield  {title} {\bibinfo
  {title} {Neural-network quantum state tomography},\ }\href
  {https://doi.org/10.1038/s41567-018-0048-5} {\bibfield  {journal} {\bibinfo
  {journal} {Nat. Phys.}\ }\textbf {\bibinfo {volume} {14}},\ \bibinfo {pages}
  {447} (\bibinfo {year} {2018})}\BibitemShut {NoStop}%
\bibitem [{\citenamefont {Carrasquilla}\ \emph {et~al.}(2019)\citenamefont
  {Carrasquilla}, \citenamefont {Torlai}, \citenamefont {Melko},\ and\
  \citenamefont {Aolita}}]{Carrasquilla2019}%
  \BibitemOpen
  \bibfield  {author} {\bibinfo {author} {\bibfnamefont {J.}~\bibnamefont
  {Carrasquilla}}, \bibinfo {author} {\bibfnamefont {G.}~\bibnamefont
  {Torlai}}, \bibinfo {author} {\bibfnamefont {R.~G.}\ \bibnamefont {Melko}},\
  and\ \bibinfo {author} {\bibfnamefont {L.}~\bibnamefont {Aolita}},\
  }\bibfield  {title} {\bibinfo {title} {Reconstructing quantum states with
  generative models},\ }\href {https://doi.org/10.1038/s42256-019-0028-1}
  {\bibfield  {journal} {\bibinfo  {journal} {Nat. Mach. Intell.}\ }\textbf
  {\bibinfo {volume} {1}},\ \bibinfo {pages} {155–161} (\bibinfo {year}
  {2019})}\BibitemShut {NoStop}%
\bibitem [{\citenamefont {Torlai}\ and\ \citenamefont
  {Melko}(2020)}]{Torlai2020}%
  \BibitemOpen
  \bibfield  {author} {\bibinfo {author} {\bibfnamefont {G.}~\bibnamefont
  {Torlai}}\ and\ \bibinfo {author} {\bibfnamefont {R.~G.}\ \bibnamefont
  {Melko}},\ }\bibfield  {title} {\bibinfo {title} {Machine-learning quantum
  states in the nisq era},\ }\href
  {https://doi.org/10.1146/annurev-conmatphys-031119-050651} {\bibfield
  {journal} {\bibinfo  {journal} {Annu. Rev. Condens. Matter Phys.}\ }\textbf
  {\bibinfo {volume} {11}},\ \bibinfo {pages} {325} (\bibinfo {year}
  {2020})}\BibitemShut {NoStop}%
\bibitem [{\citenamefont {Golubeva}\ and\ \citenamefont
  {Melko}(2022)}]{Golubeva2022}%
  \BibitemOpen
  \bibfield  {author} {\bibinfo {author} {\bibfnamefont {A.}~\bibnamefont
  {Golubeva}}\ and\ \bibinfo {author} {\bibfnamefont {R.~G.}\ \bibnamefont
  {Melko}},\ }\bibfield  {title} {\bibinfo {title} {Pruning a restricted
  boltzmann machine for quantum state reconstruction},\ }\href
  {https://doi.org/10.1103/PhysRevB.105.125124} {\bibfield  {journal} {\bibinfo
   {journal} {Phys. Rev. B}\ }\textbf {\bibinfo {volume} {105}},\ \bibinfo
  {pages} {125124} (\bibinfo {year} {2022})}\BibitemShut {NoStop}%
\bibitem [{\citenamefont {Zhu}\ \emph {et~al.}(2022)\citenamefont {Zhu},
  \citenamefont {Wu}, \citenamefont {Bai}, \citenamefont {Wang}, \citenamefont
  {Wang},\ and\ \citenamefont {Chiribella}}]{Zhu2022}%
  \BibitemOpen
  \bibfield  {author} {\bibinfo {author} {\bibfnamefont {Y.}~\bibnamefont
  {Zhu}}, \bibinfo {author} {\bibfnamefont {Y.-D.}\ \bibnamefont {Wu}},
  \bibinfo {author} {\bibfnamefont {G.}~\bibnamefont {Bai}}, \bibinfo {author}
  {\bibfnamefont {D.-S.}\ \bibnamefont {Wang}}, \bibinfo {author}
  {\bibfnamefont {Y.}~\bibnamefont {Wang}},\ and\ \bibinfo {author}
  {\bibfnamefont {G.}~\bibnamefont {Chiribella}},\ }\bibfield  {title}
  {\bibinfo {title} {Flexible learning of quantum states with generative query
  neural networks},\ }\href {https://doi.org/10.1038/s41467-022-33928-z}
  {\bibfield  {journal} {\bibinfo  {journal} {Nat. Commun.}\ }\textbf {\bibinfo
  {volume} {13}},\ \bibinfo {pages} {6222} (\bibinfo {year}
  {2022})}\BibitemShut {NoStop}%
\bibitem [{\citenamefont {Zhao}\ \emph {et~al.}(2024)\citenamefont {Zhao},
  \citenamefont {Carleo},\ and\ \citenamefont {Vicentini}}]{Zhao2024}%
  \BibitemOpen
  \bibfield  {author} {\bibinfo {author} {\bibfnamefont {H.}~\bibnamefont
  {Zhao}}, \bibinfo {author} {\bibfnamefont {G.}~\bibnamefont {Carleo}},\ and\
  \bibinfo {author} {\bibfnamefont {F.}~\bibnamefont {Vicentini}},\ }\bibfield
  {title} {\bibinfo {title} {Empirical {S}ample {C}omplexity of {N}eural
  {N}etwork {M}ixed {S}tate {R}econstruction},\ }\href
  {https://doi.org/10.22331/q-2024-05-23-1358} {\bibfield  {journal} {\bibinfo
  {journal} {Quantum}\ }\textbf {\bibinfo {volume} {8}},\ \bibinfo {pages}
  {1358} (\bibinfo {year} {2024})}\BibitemShut {NoStop}%
\bibitem [{\citenamefont {Chen}\ \emph {et~al.}(2018)\citenamefont {Chen},
  \citenamefont {Cheng}, \citenamefont {Xie}, \citenamefont {Wang},\ and\
  \citenamefont {Xiang}}]{Chen2018}%
  \BibitemOpen
  \bibfield  {author} {\bibinfo {author} {\bibfnamefont {J.}~\bibnamefont
  {Chen}}, \bibinfo {author} {\bibfnamefont {S.}~\bibnamefont {Cheng}},
  \bibinfo {author} {\bibfnamefont {H.}~\bibnamefont {Xie}}, \bibinfo {author}
  {\bibfnamefont {L.}~\bibnamefont {Wang}},\ and\ \bibinfo {author}
  {\bibfnamefont {T.}~\bibnamefont {Xiang}},\ }\bibfield  {title} {\bibinfo
  {title} {Equivalence of restricted boltzmann machines and tensor network
  states},\ }\href {https://doi.org/10.1103/PhysRevB.97.085104} {\bibfield
  {journal} {\bibinfo  {journal} {Phys. Rev. B}\ }\textbf {\bibinfo {volume}
  {97}},\ \bibinfo {pages} {085104} (\bibinfo {year} {2018})}\BibitemShut
  {NoStop}%
\bibitem [{\citenamefont {Huang}\ and\ \citenamefont
  {Moore}(2021)}]{Huang2021}%
  \BibitemOpen
  \bibfield  {author} {\bibinfo {author} {\bibfnamefont {Y.}~\bibnamefont
  {Huang}}\ and\ \bibinfo {author} {\bibfnamefont {J.~E.}\ \bibnamefont
  {Moore}},\ }\bibfield  {title} {\bibinfo {title} {Neural network
  representation of tensor network and chiral states},\ }\href
  {https://doi.org/10.1103/PhysRevLett.127.170601} {\bibfield  {journal}
  {\bibinfo  {journal} {Phys. Rev. Lett.}\ }\textbf {\bibinfo {volume} {127}},\
  \bibinfo {pages} {170601} (\bibinfo {year} {2021})}\BibitemShut {NoStop}%
\bibitem [{\citenamefont {Wang}\ \emph {et~al.}(2023)\citenamefont {Wang},
  \citenamefont {Pan}, \citenamefont {Xu}, \citenamefont {Yang}, \citenamefont
  {Li},\ and\ \citenamefont {Cichocki}}]{Wang2023}%
  \BibitemOpen
  \bibfield  {author} {\bibinfo {author} {\bibfnamefont {M.}~\bibnamefont
  {Wang}}, \bibinfo {author} {\bibfnamefont {Y.}~\bibnamefont {Pan}}, \bibinfo
  {author} {\bibfnamefont {Z.}~\bibnamefont {Xu}}, \bibinfo {author}
  {\bibfnamefont {X.}~\bibnamefont {Yang}}, \bibinfo {author} {\bibfnamefont
  {G.}~\bibnamefont {Li}},\ and\ \bibinfo {author} {\bibfnamefont
  {A.}~\bibnamefont {Cichocki}},\ }\href@noop {} {\bibinfo {title} {Tensor
  networks meet neural networks: A survey and future perspectives}} (\bibinfo
  {year} {2023}),\ \Eprint {https://arxiv.org/abs/2302.09019}
  {arXiv:2302.09019} \BibitemShut {NoStop}%
\bibitem [{\citenamefont {Zhao}\ \emph {et~al.}(2023)\citenamefont {Zhao},
  \citenamefont {Lewis}, \citenamefont {Kannan}, \citenamefont {Quek},
  \citenamefont {Huang},\ and\ \citenamefont {Caro}}]{Zhao2023}%
  \BibitemOpen
  \bibfield  {author} {\bibinfo {author} {\bibfnamefont {H.}~\bibnamefont
  {Zhao}}, \bibinfo {author} {\bibfnamefont {L.}~\bibnamefont {Lewis}},
  \bibinfo {author} {\bibfnamefont {I.}~\bibnamefont {Kannan}}, \bibinfo
  {author} {\bibfnamefont {Y.}~\bibnamefont {Quek}}, \bibinfo {author}
  {\bibfnamefont {H.-Y.}\ \bibnamefont {Huang}},\ and\ \bibinfo {author}
  {\bibfnamefont {M.~C.}\ \bibnamefont {Caro}},\ }\href@noop {} {\bibinfo
  {title} {Learning quantum states and unitaries of bounded gate complexity}}
  (\bibinfo {year} {2023}),\ \Eprint {https://arxiv.org/abs/2310.19882}
  {arXiv:2310.19882} \BibitemShut {NoStop}%
\bibitem [{\citenamefont {Noh}\ \emph {et~al.}(2020)\citenamefont {Noh},
  \citenamefont {Jiang},\ and\ \citenamefont {Fefferman}}]{Noh2020}%
  \BibitemOpen
  \bibfield  {author} {\bibinfo {author} {\bibfnamefont {K.}~\bibnamefont
  {Noh}}, \bibinfo {author} {\bibfnamefont {L.}~\bibnamefont {Jiang}},\ and\
  \bibinfo {author} {\bibfnamefont {B.}~\bibnamefont {Fefferman}},\ }\bibfield
  {title} {\bibinfo {title} {Efficient classical simulation of noisy random
  quantum circuits in one dimension},\ }\href
  {https://doi.org/10.22331/q-2020-09-11-318} {\bibfield  {journal} {\bibinfo
  {journal} {{Quantum}}\ }\textbf {\bibinfo {volume} {4}},\ \bibinfo {pages}
  {318} (\bibinfo {year} {2020})}\BibitemShut {NoStop}%
\bibitem [{\citenamefont {Aaronson}(2018)}]{Aaronson2018}%
  \BibitemOpen
  \bibfield  {author} {\bibinfo {author} {\bibfnamefont {S.}~\bibnamefont
  {Aaronson}},\ }\bibfield  {title} {\bibinfo {title} {Shadow tomography of
  quantum states},\ }in\ \href {https://doi.org/10.1145/3188745.3188802} {\emph
  {\bibinfo {booktitle} {STOC 2018}}}\ (\bibinfo {year} {2018})\ p.\ \bibinfo
  {pages} {325–338}\BibitemShut {NoStop}%
\bibitem [{\citenamefont {Huang}\ \emph {et~al.}(2020)\citenamefont {Huang},
  \citenamefont {Kueng},\ and\ \citenamefont {Preskill}}]{Huang2020}%
  \BibitemOpen
  \bibfield  {author} {\bibinfo {author} {\bibfnamefont {H.-Y.}\ \bibnamefont
  {Huang}}, \bibinfo {author} {\bibfnamefont {R.}~\bibnamefont {Kueng}},\ and\
  \bibinfo {author} {\bibfnamefont {J.}~\bibnamefont {Preskill}},\ }\bibfield
  {title} {\bibinfo {title} {Predicting many properties of a quantum system
  from very few measurements},\ }\href
  {https://doi.org/10.1038/s41567-020-0932-7} {\bibfield  {journal} {\bibinfo
  {journal} {Nat. Phys.}\ }\textbf {\bibinfo {volume} {16}},\ \bibinfo {pages}
  {1050–1057} (\bibinfo {year} {2020})}\BibitemShut {NoStop}%
\bibitem [{\citenamefont {Hu}\ and\ \citenamefont {You}(2022)}]{Hu2022}%
  \BibitemOpen
  \bibfield  {author} {\bibinfo {author} {\bibfnamefont {H.-Y.}\ \bibnamefont
  {Hu}}\ and\ \bibinfo {author} {\bibfnamefont {Y.-Z.}\ \bibnamefont {You}},\
  }\bibfield  {title} {\bibinfo {title} {Hamiltonian-driven shadow tomography
  of quantum states},\ }\href
  {https://doi.org/10.1103/PhysRevResearch.4.013054} {\bibfield  {journal}
  {\bibinfo  {journal} {Phys. Rev. Res.}\ }\textbf {\bibinfo {volume} {4}},\
  \bibinfo {pages} {013054} (\bibinfo {year} {2022})}\BibitemShut {NoStop}%
\bibitem [{\citenamefont {Liu}\ \emph {et~al.}(2023)\citenamefont {Liu},
  \citenamefont {Hao},\ and\ \citenamefont {Hu}}]{Liu2023}%
  \BibitemOpen
  \bibfield  {author} {\bibinfo {author} {\bibfnamefont {Z.}~\bibnamefont
  {Liu}}, \bibinfo {author} {\bibfnamefont {Z.}~\bibnamefont {Hao}},\ and\
  \bibinfo {author} {\bibfnamefont {H.-Y.}\ \bibnamefont {Hu}},\ }\href@noop {}
  {\bibinfo {title} {Predicting arbitrary state properties from single
  hamiltonian quench dynamics}} (\bibinfo {year} {2023}),\ \Eprint
  {https://arxiv.org/abs/2311.00695} {arXiv:2311.00695} \BibitemShut {NoStop}%
\bibitem [{\citenamefont {Kunjummen}\ \emph {et~al.}(2023)\citenamefont
  {Kunjummen}, \citenamefont {Tran}, \citenamefont {Carney},\ and\
  \citenamefont {Taylor}}]{Kunjummen2023}%
  \BibitemOpen
  \bibfield  {author} {\bibinfo {author} {\bibfnamefont {J.}~\bibnamefont
  {Kunjummen}}, \bibinfo {author} {\bibfnamefont {M.~C.}\ \bibnamefont {Tran}},
  \bibinfo {author} {\bibfnamefont {D.}~\bibnamefont {Carney}},\ and\ \bibinfo
  {author} {\bibfnamefont {J.~M.}\ \bibnamefont {Taylor}},\ }\bibfield  {title}
  {\bibinfo {title} {Shadow process tomography of quantum channels},\ }\href
  {https://doi.org/10.1103/PhysRevA.107.042403} {\bibfield  {journal} {\bibinfo
   {journal} {Phys. Rev. A}\ }\textbf {\bibinfo {volume} {107}},\ \bibinfo
  {pages} {042403} (\bibinfo {year} {2023})}\BibitemShut {NoStop}%
\bibitem [{\citenamefont {Levy}\ \emph {et~al.}(2024)\citenamefont {Levy},
  \citenamefont {Luo},\ and\ \citenamefont {Clark}}]{Levy2024}%
  \BibitemOpen
  \bibfield  {author} {\bibinfo {author} {\bibfnamefont {R.}~\bibnamefont
  {Levy}}, \bibinfo {author} {\bibfnamefont {D.}~\bibnamefont {Luo}},\ and\
  \bibinfo {author} {\bibfnamefont {B.~K.}\ \bibnamefont {Clark}},\ }\bibfield
  {title} {\bibinfo {title} {Classical shadows for quantum process tomography
  on near-term quantum computers},\ }\href
  {https://doi.org/10.1103/PhysRevResearch.6.013029} {\bibfield  {journal}
  {\bibinfo  {journal} {Phys. Rev. Res.}\ }\textbf {\bibinfo {volume} {6}},\
  \bibinfo {pages} {013029} (\bibinfo {year} {2024})}\BibitemShut {NoStop}%
\bibitem [{\citenamefont {Verstraete}\ \emph {et~al.}(2004)\citenamefont
  {Verstraete}, \citenamefont {Garc\'{\i}a-Ripoll},\ and\ \citenamefont
  {Cirac}}]{Verstraete2004A}%
  \BibitemOpen
  \bibfield  {author} {\bibinfo {author} {\bibfnamefont {F.}~\bibnamefont
  {Verstraete}}, \bibinfo {author} {\bibfnamefont {J.~J.}\ \bibnamefont
  {Garc\'{\i}a-Ripoll}},\ and\ \bibinfo {author} {\bibfnamefont {J.~I.}\
  \bibnamefont {Cirac}},\ }\bibfield  {title} {\bibinfo {title} {Matrix product
  density operators: Simulation of finite-temperature and dissipative
  systems},\ }\href {https://doi.org/10.1103/PhysRevLett.93.207204} {\bibfield
  {journal} {\bibinfo  {journal} {Phys. Rev. Lett.}\ }\textbf {\bibinfo
  {volume} {93}},\ \bibinfo {pages} {207204} (\bibinfo {year}
  {2004})}\BibitemShut {NoStop}%
\bibitem [{\citenamefont {Verstraete}\ \emph {et~al.}(2006)\citenamefont
  {Verstraete}, \citenamefont {Wolf}, \citenamefont {Perez-Garcia},\ and\
  \citenamefont {Cirac}}]{Verstraete2006B}%
  \BibitemOpen
  \bibfield  {author} {\bibinfo {author} {\bibfnamefont {F.}~\bibnamefont
  {Verstraete}}, \bibinfo {author} {\bibfnamefont {M.~M.}\ \bibnamefont
  {Wolf}}, \bibinfo {author} {\bibfnamefont {D.}~\bibnamefont {Perez-Garcia}},\
  and\ \bibinfo {author} {\bibfnamefont {J.~I.}\ \bibnamefont {Cirac}},\
  }\bibfield  {title} {\bibinfo {title} {Criticality, the area law, and the
  computational power of projected entangled pair states},\ }\href
  {https://doi.org/10.1103/PhysRevLett.96.220601} {\bibfield  {journal}
  {\bibinfo  {journal} {Phys. Rev. Lett.}\ }\textbf {\bibinfo {volume} {96}},\
  \bibinfo {pages} {220601} (\bibinfo {year} {2006})}\BibitemShut {NoStop}%
\bibitem [{\citenamefont {Schuch}\ \emph {et~al.}(2007)\citenamefont {Schuch},
  \citenamefont {Wolf}, \citenamefont {Verstraete},\ and\ \citenamefont
  {Cirac}}]{Schuch2007}%
  \BibitemOpen
  \bibfield  {author} {\bibinfo {author} {\bibfnamefont {N.}~\bibnamefont
  {Schuch}}, \bibinfo {author} {\bibfnamefont {M.~M.}\ \bibnamefont {Wolf}},
  \bibinfo {author} {\bibfnamefont {F.}~\bibnamefont {Verstraete}},\ and\
  \bibinfo {author} {\bibfnamefont {J.~I.}\ \bibnamefont {Cirac}},\ }\bibfield
  {title} {\bibinfo {title} {Computational complexity of projected entangled
  pair states},\ }\href {https://doi.org/10.1103/PhysRevLett.98.140506}
  {\bibfield  {journal} {\bibinfo  {journal} {Phys. Rev. Lett.}\ }\textbf
  {\bibinfo {volume} {98}},\ \bibinfo {pages} {140506} (\bibinfo {year}
  {2007})}\BibitemShut {NoStop}%
\bibitem [{\citenamefont {P\'erez-Garc\'{\i}a}\ \emph
  {et~al.}(2008)\citenamefont {P\'erez-Garc\'{\i}a}, \citenamefont
  {Verstraete}, \citenamefont {Wolf},\ and\ \citenamefont {Cirac}}]{Perez2008}%
  \BibitemOpen
  \bibfield  {author} {\bibinfo {author} {\bibfnamefont {D.}~\bibnamefont
  {P\'erez-Garc\'{\i}a}}, \bibinfo {author} {\bibfnamefont {F.}~\bibnamefont
  {Verstraete}}, \bibinfo {author} {\bibfnamefont {M.~M.}\ \bibnamefont
  {Wolf}},\ and\ \bibinfo {author} {\bibfnamefont {J.~I.}\ \bibnamefont
  {Cirac}},\ }\bibfield  {title} {\bibinfo {title} {Peps as unique ground
  states of local hamiltonians},\ }\href {https://doi.org/10.26421/QIC8.6-7-6}
  {\bibfield  {journal} {\bibinfo  {journal} {Quantum Info. Comput.}\ }\textbf
  {\bibinfo {volume} {8}},\ \bibinfo {pages} {650–663} (\bibinfo {year}
  {2008})}\BibitemShut {NoStop}%
\bibitem [{\citenamefont {Zaletel}\ and\ \citenamefont
  {Pollmann}(2020)}]{Zaletel2020}%
  \BibitemOpen
  \bibfield  {author} {\bibinfo {author} {\bibfnamefont {M.~P.}\ \bibnamefont
  {Zaletel}}\ and\ \bibinfo {author} {\bibfnamefont {F.}~\bibnamefont
  {Pollmann}},\ }\bibfield  {title} {\bibinfo {title} {Isometric tensor network
  states in two dimensions},\ }\href
  {https://doi.org/10.1103/PhysRevLett.124.037201} {\bibfield  {journal}
  {\bibinfo  {journal} {Phys. Rev. Lett.}\ }\textbf {\bibinfo {volume} {124}},\
  \bibinfo {pages} {037201} (\bibinfo {year} {2020})}\BibitemShut {NoStop}%
\bibitem [{\citenamefont {Soejima}\ \emph {et~al.}(2020)\citenamefont
  {Soejima}, \citenamefont {Siva}, \citenamefont {Bultinck}, \citenamefont
  {Chatterjee}, \citenamefont {Pollmann},\ and\ \citenamefont
  {Zaletel}}]{Soejima2020}%
  \BibitemOpen
  \bibfield  {author} {\bibinfo {author} {\bibfnamefont {T.}~\bibnamefont
  {Soejima}}, \bibinfo {author} {\bibfnamefont {K.}~\bibnamefont {Siva}},
  \bibinfo {author} {\bibfnamefont {N.}~\bibnamefont {Bultinck}}, \bibinfo
  {author} {\bibfnamefont {S.}~\bibnamefont {Chatterjee}}, \bibinfo {author}
  {\bibfnamefont {F.}~\bibnamefont {Pollmann}},\ and\ \bibinfo {author}
  {\bibfnamefont {M.~P.}\ \bibnamefont {Zaletel}},\ }\bibfield  {title}
  {\bibinfo {title} {Isometric tensor network representation of string-net
  liquids},\ }\href {https://doi.org/10.1103/PhysRevB.101.085117} {\bibfield
  {journal} {\bibinfo  {journal} {Phys. Rev. B}\ }\textbf {\bibinfo {volume}
  {101}},\ \bibinfo {pages} {085117} (\bibinfo {year} {2020})}\BibitemShut
  {NoStop}%
\bibitem [{\citenamefont {Kadow}\ \emph {et~al.}(2023)\citenamefont {Kadow},
  \citenamefont {Pollmann},\ and\ \citenamefont {Knap}}]{Kadow2023}%
  \BibitemOpen
  \bibfield  {author} {\bibinfo {author} {\bibfnamefont {W.}~\bibnamefont
  {Kadow}}, \bibinfo {author} {\bibfnamefont {F.}~\bibnamefont {Pollmann}},\
  and\ \bibinfo {author} {\bibfnamefont {M.}~\bibnamefont {Knap}},\ }\bibfield
  {title} {\bibinfo {title} {Isometric tensor network representations of
  two-dimensional thermal states},\ }\href
  {https://doi.org/10.1103/PhysRevB.107.205106} {\bibfield  {journal} {\bibinfo
   {journal} {Phys. Rev. B}\ }\textbf {\bibinfo {volume} {107}},\ \bibinfo
  {pages} {205106} (\bibinfo {year} {2023})}\BibitemShut {NoStop}%
\bibitem [{\citenamefont {Liu}\ \emph {et~al.}(2024)\citenamefont {Liu},
  \citenamefont {Shtengel},\ and\ \citenamefont {Pollmann}}]{Liu2024}%
  \BibitemOpen
  \bibfield  {author} {\bibinfo {author} {\bibfnamefont {Y.-J.}\ \bibnamefont
  {Liu}}, \bibinfo {author} {\bibfnamefont {K.}~\bibnamefont {Shtengel}},\ and\
  \bibinfo {author} {\bibfnamefont {F.}~\bibnamefont {Pollmann}},\ }\href@noop
  {} {\bibinfo {title} {Topological quantum phase transitions in 2d isometric
  tensor networks}} (\bibinfo {year} {2024}),\ \Eprint
  {https://arxiv.org/abs/2312.05079} {arXiv:2312.05079} \BibitemShut {NoStop}%
\bibitem [{\citenamefont {Werner}\ \emph {et~al.}(2016)\citenamefont {Werner},
  \citenamefont {Jaschke}, \citenamefont {Silvi}, \citenamefont {Kliesch},
  \citenamefont {Calarco}, \citenamefont {Eisert},\ and\ \citenamefont
  {Montangero}}]{Werner2016}%
  \BibitemOpen
  \bibfield  {author} {\bibinfo {author} {\bibfnamefont {A.~H.}\ \bibnamefont
  {Werner}}, \bibinfo {author} {\bibfnamefont {D.}~\bibnamefont {Jaschke}},
  \bibinfo {author} {\bibfnamefont {P.}~\bibnamefont {Silvi}}, \bibinfo
  {author} {\bibfnamefont {M.}~\bibnamefont {Kliesch}}, \bibinfo {author}
  {\bibfnamefont {T.}~\bibnamefont {Calarco}}, \bibinfo {author} {\bibfnamefont
  {J.}~\bibnamefont {Eisert}},\ and\ \bibinfo {author} {\bibfnamefont
  {S.}~\bibnamefont {Montangero}},\ }\bibfield  {title} {\bibinfo {title}
  {Positive tensor network approach for simulating open quantum many-body
  systems},\ }\href {https://doi.org/10.1103/PhysRevLett.116.237201} {\bibfield
   {journal} {\bibinfo  {journal} {Phys. Rev. Lett.}\ }\textbf {\bibinfo
  {volume} {116}},\ \bibinfo {pages} {237201} (\bibinfo {year}
  {2016})}\BibitemShut {NoStop}%
\bibitem [{\citenamefont {Cheng}\ \emph {et~al.}(2021)\citenamefont {Cheng},
  \citenamefont {Cao}, \citenamefont {Zhang}, \citenamefont {Liu},
  \citenamefont {Hou}, \citenamefont {Xu},\ and\ \citenamefont
  {Zeng}}]{Cheng2021}%
  \BibitemOpen
  \bibfield  {author} {\bibinfo {author} {\bibfnamefont {S.}~\bibnamefont
  {Cheng}}, \bibinfo {author} {\bibfnamefont {C.}~\bibnamefont {Cao}}, \bibinfo
  {author} {\bibfnamefont {C.}~\bibnamefont {Zhang}}, \bibinfo {author}
  {\bibfnamefont {Y.}~\bibnamefont {Liu}}, \bibinfo {author} {\bibfnamefont
  {S.-Y.}\ \bibnamefont {Hou}}, \bibinfo {author} {\bibfnamefont
  {P.}~\bibnamefont {Xu}},\ and\ \bibinfo {author} {\bibfnamefont
  {B.}~\bibnamefont {Zeng}},\ }\bibfield  {title} {\bibinfo {title} {Simulating
  noisy quantum circuits with matrix product density operators},\ }\href
  {https://doi.org/10.1103/PhysRevResearch.3.023005} {\bibfield  {journal}
  {\bibinfo  {journal} {Phys. Rev. Res.}\ }\textbf {\bibinfo {volume} {3}},\
  \bibinfo {pages} {023005} (\bibinfo {year} {2021})}\BibitemShut {NoStop}%
\bibitem [{\citenamefont {Guo}\ and\ \citenamefont {Yang}(2024)}]{Guo2023C}%
  \BibitemOpen
  \bibfield  {author} {\bibinfo {author} {\bibfnamefont {Y.}~\bibnamefont
  {Guo}}\ and\ \bibinfo {author} {\bibfnamefont {S.}~\bibnamefont {Yang}},\
  }\href@noop {} {\bibinfo {title} {Locally purified density operators for
  noisy quantum circuits}} (\bibinfo {year} {2024}),\ \Eprint
  {https://arxiv.org/abs/2312.02854} {arXiv:2312.02854} \BibitemShut {NoStop}%
\bibitem [{\citenamefont {Guo}\ \emph {et~al.}(2024)\citenamefont {Guo},
  \citenamefont {Zhang}, \citenamefont {Yang},\ and\ \citenamefont
  {Bi}}]{Guo2024}%
  \BibitemOpen
  \bibfield  {author} {\bibinfo {author} {\bibfnamefont {Y.}~\bibnamefont
  {Guo}}, \bibinfo {author} {\bibfnamefont {J.-H.}\ \bibnamefont {Zhang}},
  \bibinfo {author} {\bibfnamefont {S.}~\bibnamefont {Yang}},\ and\ \bibinfo
  {author} {\bibfnamefont {Z.}~\bibnamefont {Bi}},\ }\href@noop {} {\bibinfo
  {title} {Locally purified density operators for symmetry-protected
  topological phases in mixed states}} (\bibinfo {year} {2024}),\ \Eprint
  {https://arxiv.org/abs/2403.16978} {arXiv:2403.16978} \BibitemShut {NoStop}%
\bibitem [{\citenamefont {Kingma}\ and\ \citenamefont {Ba}(2017)}]{Kingma2017}%
  \BibitemOpen
  \bibfield  {author} {\bibinfo {author} {\bibfnamefont {D.~P.}\ \bibnamefont
  {Kingma}}\ and\ \bibinfo {author} {\bibfnamefont {J.}~\bibnamefont {Ba}},\
  }\href@noop {} {\bibinfo {title} {Adam: A method for stochastic
  optimization}} (\bibinfo {year} {2017}),\ \Eprint
  {https://arxiv.org/abs/1412.6980} {arXiv:1412.6980} \BibitemShut {NoStop}%
\bibitem [{\citenamefont {Crosswhite}\ and\ \citenamefont
  {Bacon}(2008)}]{Crosswhite2008}%
  \BibitemOpen
  \bibfield  {author} {\bibinfo {author} {\bibfnamefont {G.~M.}\ \bibnamefont
  {Crosswhite}}\ and\ \bibinfo {author} {\bibfnamefont {D.}~\bibnamefont
  {Bacon}},\ }\bibfield  {title} {\bibinfo {title} {Finite automata for caching
  in matrix product algorithms},\ }\href
  {https://doi.org/10.1103/PhysRevA.78.012356} {\bibfield  {journal} {\bibinfo
  {journal} {Phys. Rev. A}\ }\textbf {\bibinfo {volume} {78}},\ \bibinfo
  {pages} {012356} (\bibinfo {year} {2008})}\BibitemShut {NoStop}%
\bibitem [{\citenamefont {Verstraete}\ and\ \citenamefont
  {Cirac}(2004)}]{verstraete2004B}%
  \BibitemOpen
  \bibfield  {author} {\bibinfo {author} {\bibfnamefont {F.}~\bibnamefont
  {Verstraete}}\ and\ \bibinfo {author} {\bibfnamefont {J.~I.}\ \bibnamefont
  {Cirac}},\ }\href@noop {} {\bibinfo {title} {Renormalization algorithms for
  quantum-many body systems in two and higher dimensions}} (\bibinfo {year}
  {2004}),\ \Eprint {https://arxiv.org/abs/cond-mat/0407066}
  {arXiv:cond-mat/0407066} \BibitemShut {NoStop}%
\bibitem [{\citenamefont {Slattery}\ and\ \citenamefont
  {Clark}(2021)}]{Slattery2021}%
  \BibitemOpen
  \bibfield  {author} {\bibinfo {author} {\bibfnamefont {L.}~\bibnamefont
  {Slattery}}\ and\ \bibinfo {author} {\bibfnamefont {B.~K.}\ \bibnamefont
  {Clark}},\ }\href@noop {} {\bibinfo {title} {Quantum circuits for
  two-dimensional isometric tensor networks}} (\bibinfo {year} {2021}),\
  \Eprint {https://arxiv.org/abs/2108.02792} {arXiv:2108.02792} \BibitemShut
  {NoStop}%
\bibitem [{\citenamefont {Lin}\ \emph {et~al.}(2022)\citenamefont {Lin},
  \citenamefont {Zaletel},\ and\ \citenamefont {Pollmann}}]{Lin2022}%
  \BibitemOpen
  \bibfield  {author} {\bibinfo {author} {\bibfnamefont {S.-H.}\ \bibnamefont
  {Lin}}, \bibinfo {author} {\bibfnamefont {M.~P.}\ \bibnamefont {Zaletel}},\
  and\ \bibinfo {author} {\bibfnamefont {F.}~\bibnamefont {Pollmann}},\
  }\bibfield  {title} {\bibinfo {title} {Efficient simulation of dynamics in
  two-dimensional quantum spin systems with isometric tensor networks},\ }\href
  {https://doi.org/10.1103/PhysRevB.106.245102} {\bibfield  {journal} {\bibinfo
   {journal} {Phys. Rev. B}\ }\textbf {\bibinfo {volume} {106}},\ \bibinfo
  {pages} {245102} (\bibinfo {year} {2022})}\BibitemShut {NoStop}%
\bibitem [{\citenamefont {Gu}\ and\ \citenamefont {Wen}(2009)}]{Gu2009}%
  \BibitemOpen
  \bibfield  {author} {\bibinfo {author} {\bibfnamefont {Z.-C.}\ \bibnamefont
  {Gu}}\ and\ \bibinfo {author} {\bibfnamefont {X.-G.}\ \bibnamefont {Wen}},\
  }\bibfield  {title} {\bibinfo {title} {Tensor-entanglement-filtering
  renormalization approach and symmetry-protected topological order},\ }\href
  {https://doi.org/10.1103/PhysRevB.80.155131} {\bibfield  {journal} {\bibinfo
  {journal} {Phys. Rev. B}\ }\textbf {\bibinfo {volume} {80}},\ \bibinfo
  {pages} {155131} (\bibinfo {year} {2009})}\BibitemShut {NoStop}%
\bibitem [{\citenamefont {Chen}\ \emph {et~al.}(2010)\citenamefont {Chen},
  \citenamefont {Gu},\ and\ \citenamefont {Wen}}]{Chen2010}%
  \BibitemOpen
  \bibfield  {author} {\bibinfo {author} {\bibfnamefont {X.}~\bibnamefont
  {Chen}}, \bibinfo {author} {\bibfnamefont {Z.-C.}\ \bibnamefont {Gu}},\ and\
  \bibinfo {author} {\bibfnamefont {X.-G.}\ \bibnamefont {Wen}},\ }\bibfield
  {title} {\bibinfo {title} {Local unitary transformation, long-range quantum
  entanglement, wave function renormalization, and topological order},\ }\href
  {https://doi.org/10.1103/PhysRevB.82.155138} {\bibfield  {journal} {\bibinfo
  {journal} {Phys. Rev. B}\ }\textbf {\bibinfo {volume} {82}},\ \bibinfo
  {pages} {155138} (\bibinfo {year} {2010})}\BibitemShut {NoStop}%
\bibitem [{\citenamefont {Pollmann}\ \emph {et~al.}(2010)\citenamefont
  {Pollmann}, \citenamefont {Turner}, \citenamefont {Berg},\ and\ \citenamefont
  {Oshikawa}}]{Pollmann2010}%
  \BibitemOpen
  \bibfield  {author} {\bibinfo {author} {\bibfnamefont {F.}~\bibnamefont
  {Pollmann}}, \bibinfo {author} {\bibfnamefont {A.~M.}\ \bibnamefont
  {Turner}}, \bibinfo {author} {\bibfnamefont {E.}~\bibnamefont {Berg}},\ and\
  \bibinfo {author} {\bibfnamefont {M.}~\bibnamefont {Oshikawa}},\ }\bibfield
  {title} {\bibinfo {title} {Entanglement spectrum of a topological phase in
  one dimension},\ }\href {https://doi.org/10.1103/PhysRevB.81.064439}
  {\bibfield  {journal} {\bibinfo  {journal} {Phys. Rev. B}\ }\textbf {\bibinfo
  {volume} {81}},\ \bibinfo {pages} {064439} (\bibinfo {year}
  {2010})}\BibitemShut {NoStop}%
\bibitem [{\citenamefont {Chen}\ \emph {et~al.}(2011)\citenamefont {Chen},
  \citenamefont {Gu},\ and\ \citenamefont {Wen}}]{Chen2011}%
  \BibitemOpen
  \bibfield  {author} {\bibinfo {author} {\bibfnamefont {X.}~\bibnamefont
  {Chen}}, \bibinfo {author} {\bibfnamefont {Z.-C.}\ \bibnamefont {Gu}},\ and\
  \bibinfo {author} {\bibfnamefont {X.-G.}\ \bibnamefont {Wen}},\ }\bibfield
  {title} {\bibinfo {title} {Complete classification of one-dimensional gapped
  quantum phases in interacting spin systems},\ }\href
  {https://doi.org/10.1103/PhysRevB.84.235128} {\bibfield  {journal} {\bibinfo
  {journal} {Phys. Rev. B}\ }\textbf {\bibinfo {volume} {84}},\ \bibinfo
  {pages} {235128} (\bibinfo {year} {2011})}\BibitemShut {NoStop}%
\bibitem [{\citenamefont {Pollmann}\ and\ \citenamefont
  {Turner}(2012)}]{Pollmann2012}%
  \BibitemOpen
  \bibfield  {author} {\bibinfo {author} {\bibfnamefont {F.}~\bibnamefont
  {Pollmann}}\ and\ \bibinfo {author} {\bibfnamefont {A.~M.}\ \bibnamefont
  {Turner}},\ }\bibfield  {title} {\bibinfo {title} {Detection of
  symmetry-protected topological phases in one dimension},\ }\href
  {https://doi.org/10.1103/PhysRevB.86.125441} {\bibfield  {journal} {\bibinfo
  {journal} {Phys. Rev. B}\ }\textbf {\bibinfo {volume} {86}},\ \bibinfo
  {pages} {125441} (\bibinfo {year} {2012})}\BibitemShut {NoStop}%
\bibitem [{\citenamefont {Chen}\ \emph {et~al.}(2013)\citenamefont {Chen},
  \citenamefont {Gu}, \citenamefont {Liu},\ and\ \citenamefont
  {Wen}}]{Chen2013}%
  \BibitemOpen
  \bibfield  {author} {\bibinfo {author} {\bibfnamefont {X.}~\bibnamefont
  {Chen}}, \bibinfo {author} {\bibfnamefont {Z.-C.}\ \bibnamefont {Gu}},
  \bibinfo {author} {\bibfnamefont {Z.-X.}\ \bibnamefont {Liu}},\ and\ \bibinfo
  {author} {\bibfnamefont {X.-G.}\ \bibnamefont {Wen}},\ }\bibfield  {title}
  {\bibinfo {title} {Symmetry protected topological orders and the group
  cohomology of their symmetry group},\ }\href
  {https://doi.org/10.1103/PhysRevB.87.155114} {\bibfield  {journal} {\bibinfo
  {journal} {Phys. Rev. B}\ }\textbf {\bibinfo {volume} {87}},\ \bibinfo
  {pages} {155114} (\bibinfo {year} {2013})}\BibitemShut {NoStop}%
\bibitem [{\citenamefont {Briegel}\ and\ \citenamefont
  {Raussendorf}(2001)}]{Briegel2001}%
  \BibitemOpen
  \bibfield  {author} {\bibinfo {author} {\bibfnamefont {H.~J.}\ \bibnamefont
  {Briegel}}\ and\ \bibinfo {author} {\bibfnamefont {R.}~\bibnamefont
  {Raussendorf}},\ }\bibfield  {title} {\bibinfo {title} {Persistent
  entanglement in arrays of interacting particles},\ }\href
  {https://doi.org/10.1103/PhysRevLett.86.910} {\bibfield  {journal} {\bibinfo
  {journal} {Phys. Rev. Lett.}\ }\textbf {\bibinfo {volume} {86}},\ \bibinfo
  {pages} {910} (\bibinfo {year} {2001})}\BibitemShut {NoStop}%
\bibitem [{\citenamefont {Raussendorf}\ and\ \citenamefont
  {Briegel}(2001)}]{Raussendorf2001}%
  \BibitemOpen
  \bibfield  {author} {\bibinfo {author} {\bibfnamefont {R.}~\bibnamefont
  {Raussendorf}}\ and\ \bibinfo {author} {\bibfnamefont {H.~J.}\ \bibnamefont
  {Briegel}},\ }\bibfield  {title} {\bibinfo {title} {A one-way quantum
  computer},\ }\href {https://doi.org/10.1103/PhysRevLett.86.5188} {\bibfield
  {journal} {\bibinfo  {journal} {Phys. Rev. Lett.}\ }\textbf {\bibinfo
  {volume} {86}},\ \bibinfo {pages} {5188} (\bibinfo {year}
  {2001})}\BibitemShut {NoStop}%
\bibitem [{\citenamefont {Briegel}\ \emph {et~al.}(2009)\citenamefont
  {Briegel}, \citenamefont {Browne}, \citenamefont {Dür}, \citenamefont
  {Raussendorf},\ and\ \citenamefont {Van~den Nest}}]{Briegel2009}%
  \BibitemOpen
  \bibfield  {author} {\bibinfo {author} {\bibfnamefont {H.~J.}\ \bibnamefont
  {Briegel}}, \bibinfo {author} {\bibfnamefont {D.~E.}\ \bibnamefont {Browne}},
  \bibinfo {author} {\bibfnamefont {W.}~\bibnamefont {Dür}}, \bibinfo {author}
  {\bibfnamefont {R.}~\bibnamefont {Raussendorf}},\ and\ \bibinfo {author}
  {\bibfnamefont {M.}~\bibnamefont {Van~den Nest}},\ }\bibfield  {title}
  {\bibinfo {title} {Measurement-based quantum computation},\ }\href
  {https://doi.org/10.1038/nphys1157} {\bibfield  {journal} {\bibinfo
  {journal} {Nat. Phys.}\ }\textbf {\bibinfo {volume} {5}},\ \bibinfo {pages}
  {19} (\bibinfo {year} {2009})}\BibitemShut {NoStop}%
\bibitem [{\citenamefont {Raussendorf}\ \emph {et~al.}(2017)\citenamefont
  {Raussendorf}, \citenamefont {Wang}, \citenamefont {Prakash}, \citenamefont
  {Wei},\ and\ \citenamefont {Stephen}}]{Raussendorf2017}%
  \BibitemOpen
  \bibfield  {author} {\bibinfo {author} {\bibfnamefont {R.}~\bibnamefont
  {Raussendorf}}, \bibinfo {author} {\bibfnamefont {D.-S.}\ \bibnamefont
  {Wang}}, \bibinfo {author} {\bibfnamefont {A.}~\bibnamefont {Prakash}},
  \bibinfo {author} {\bibfnamefont {T.-C.}\ \bibnamefont {Wei}},\ and\ \bibinfo
  {author} {\bibfnamefont {D.~T.}\ \bibnamefont {Stephen}},\ }\bibfield
  {title} {\bibinfo {title} {Symmetry-protected topological phases with uniform
  computational power in one dimension},\ }\href
  {https://doi.org/10.1103/PhysRevA.96.012302} {\bibfield  {journal} {\bibinfo
  {journal} {Phys. Rev. A}\ }\textbf {\bibinfo {volume} {96}},\ \bibinfo
  {pages} {012302} (\bibinfo {year} {2017})}\BibitemShut {NoStop}%
\bibitem [{\citenamefont {Raussendorf}\ \emph {et~al.}(2019)\citenamefont
  {Raussendorf}, \citenamefont {Okay}, \citenamefont {Wang}, \citenamefont
  {Stephen},\ and\ \citenamefont {Nautrup}}]{Raussendorf2019}%
  \BibitemOpen
  \bibfield  {author} {\bibinfo {author} {\bibfnamefont {R.}~\bibnamefont
  {Raussendorf}}, \bibinfo {author} {\bibfnamefont {C.}~\bibnamefont {Okay}},
  \bibinfo {author} {\bibfnamefont {D.-S.}\ \bibnamefont {Wang}}, \bibinfo
  {author} {\bibfnamefont {D.~T.}\ \bibnamefont {Stephen}},\ and\ \bibinfo
  {author} {\bibfnamefont {H.~P.}\ \bibnamefont {Nautrup}},\ }\bibfield
  {title} {\bibinfo {title} {Computationally universal phase of quantum
  matter},\ }\href {https://doi.org/10.1103/PhysRevLett.122.090501} {\bibfield
  {journal} {\bibinfo  {journal} {Phys. Rev. Lett.}\ }\textbf {\bibinfo
  {volume} {122}},\ \bibinfo {pages} {090501} (\bibinfo {year}
  {2019})}\BibitemShut {NoStop}%
\bibitem [{\citenamefont {Liu}\ \emph {et~al.}(2022)\citenamefont {Liu},
  \citenamefont {Zhou},\ and\ \citenamefont {Chen}}]{Liu2022}%
  \BibitemOpen
  \bibfield  {author} {\bibinfo {author} {\bibfnamefont {H.}~\bibnamefont
  {Liu}}, \bibinfo {author} {\bibfnamefont {T.}~\bibnamefont {Zhou}},\ and\
  \bibinfo {author} {\bibfnamefont {X.}~\bibnamefont {Chen}},\ }\bibfield
  {title} {\bibinfo {title} {Measurement-induced entanglement transition in a
  two-dimensional shallow circuit},\ }\href
  {https://doi.org/10.1103/PhysRevB.106.144311} {\bibfield  {journal} {\bibinfo
   {journal} {Phys. Rev. B}\ }\textbf {\bibinfo {volume} {106}},\ \bibinfo
  {pages} {144311} (\bibinfo {year} {2022})}\BibitemShut {NoStop}%
\bibitem [{\citenamefont {Guo}\ \emph {et~al.}(2023)\citenamefont {Guo},
  \citenamefont {Zhang}, \citenamefont {Bi},\ and\ \citenamefont
  {Yang}}]{Guo2023B}%
  \BibitemOpen
  \bibfield  {author} {\bibinfo {author} {\bibfnamefont {Y.}~\bibnamefont
  {Guo}}, \bibinfo {author} {\bibfnamefont {J.-H.}\ \bibnamefont {Zhang}},
  \bibinfo {author} {\bibfnamefont {Z.}~\bibnamefont {Bi}},\ and\ \bibinfo
  {author} {\bibfnamefont {S.}~\bibnamefont {Yang}},\ }\bibfield  {title}
  {\bibinfo {title} {Triggering boundary phase transitions through bulk
  measurements in two-dimensional cluster states},\ }\href
  {https://doi.org/10.1103/PhysRevResearch.5.043069} {\bibfield  {journal}
  {\bibinfo  {journal} {Phys. Rev. Res.}\ }\textbf {\bibinfo {volume} {5}},\
  \bibinfo {pages} {043069} (\bibinfo {year} {2023})}\BibitemShut {NoStop}%
\bibitem [{\citenamefont {Garnerone}\ \emph {et~al.}(2010)\citenamefont
  {Garnerone}, \citenamefont {de~Oliveira},\ and\ \citenamefont
  {Zanardi}}]{Garnerone2010}%
  \BibitemOpen
  \bibfield  {author} {\bibinfo {author} {\bibfnamefont {S.}~\bibnamefont
  {Garnerone}}, \bibinfo {author} {\bibfnamefont {T.~R.}\ \bibnamefont
  {de~Oliveira}},\ and\ \bibinfo {author} {\bibfnamefont {P.}~\bibnamefont
  {Zanardi}},\ }\bibfield  {title} {\bibinfo {title} {Typicality in random
  matrix product states},\ }\href {https://doi.org/10.1103/PhysRevA.81.032336}
  {\bibfield  {journal} {\bibinfo  {journal} {Phys. Rev. A}\ }\textbf {\bibinfo
  {volume} {81}},\ \bibinfo {pages} {032336} (\bibinfo {year}
  {2010})}\BibitemShut {NoStop}%
\bibitem [{\citenamefont {Chen}\ \emph {et~al.}(2022)\citenamefont {Chen},
  \citenamefont {Shi}, \citenamefont {Xiang}, \citenamefont {Wang},
  \citenamefont {Li}, \citenamefont {Sun}, \citenamefont {He}, \citenamefont
  {Song}, \citenamefont {Zhao}, \citenamefont {Zheng}, \citenamefont {Xu},\
  and\ \citenamefont {Fan}}]{Chen2022}%
  \BibitemOpen
  \bibfield  {author} {\bibinfo {author} {\bibfnamefont {C.-T.}\ \bibnamefont
  {Chen}}, \bibinfo {author} {\bibfnamefont {Y.-H.}\ \bibnamefont {Shi}},
  \bibinfo {author} {\bibfnamefont {Z.}~\bibnamefont {Xiang}}, \bibinfo
  {author} {\bibfnamefont {Z.-A.}\ \bibnamefont {Wang}}, \bibinfo {author}
  {\bibfnamefont {T.-M.}\ \bibnamefont {Li}}, \bibinfo {author} {\bibfnamefont
  {H.-Y.}\ \bibnamefont {Sun}}, \bibinfo {author} {\bibfnamefont {T.-S.}\
  \bibnamefont {He}}, \bibinfo {author} {\bibfnamefont {X.}~\bibnamefont
  {Song}}, \bibinfo {author} {\bibfnamefont {S.}~\bibnamefont {Zhao}}, \bibinfo
  {author} {\bibfnamefont {D.}~\bibnamefont {Zheng}}, \bibinfo {author}
  {\bibfnamefont {K.}~\bibnamefont {Xu}},\ and\ \bibinfo {author}
  {\bibfnamefont {H.}~\bibnamefont {Fan}},\ }\bibfield  {title} {\bibinfo
  {title} {Scq cloud quantum computation for generating
  greenberger-horne-zeilinger states of up to 10 qubits},\ }\href
  {https://doi.org/10.1007/s11433-022-1972-1} {\bibfield  {journal} {\bibinfo
  {journal} {Sci. China-Phys. Mech. Astron.}\ }\textbf {\bibinfo {volume}
  {65}},\ \bibinfo {pages} {110362} (\bibinfo {year} {2022})}\BibitemShut
  {NoStop}%
\bibitem [{\citenamefont {Aolita}\ \emph {et~al.}(2015)\citenamefont {Aolita},
  \citenamefont {Gogolin}, \citenamefont {Kliesch},\ and\ \citenamefont
  {Eisert}}]{Aolita2015}%
  \BibitemOpen
  \bibfield  {author} {\bibinfo {author} {\bibfnamefont {L.}~\bibnamefont
  {Aolita}}, \bibinfo {author} {\bibfnamefont {C.}~\bibnamefont {Gogolin}},
  \bibinfo {author} {\bibfnamefont {M.}~\bibnamefont {Kliesch}},\ and\ \bibinfo
  {author} {\bibfnamefont {J.}~\bibnamefont {Eisert}},\ }\bibfield  {title}
  {\bibinfo {title} {Reliable quantum certification of photonic state
  preparations},\ }\href {https://doi.org/10.1038/ncomms9498} {\bibfield
  {journal} {\bibinfo  {journal} {Nat. Commun.}\ }\textbf {\bibinfo {volume}
  {6}},\ \bibinfo {pages} {8498} (\bibinfo {year} {2015})}\BibitemShut
  {NoStop}%
\bibitem [{\citenamefont {Gluza}\ \emph {et~al.}(2018)\citenamefont {Gluza},
  \citenamefont {Kliesch}, \citenamefont {Eisert},\ and\ \citenamefont
  {Aolita}}]{Gluza2018}%
  \BibitemOpen
  \bibfield  {author} {\bibinfo {author} {\bibfnamefont {M.}~\bibnamefont
  {Gluza}}, \bibinfo {author} {\bibfnamefont {M.}~\bibnamefont {Kliesch}},
  \bibinfo {author} {\bibfnamefont {J.}~\bibnamefont {Eisert}},\ and\ \bibinfo
  {author} {\bibfnamefont {L.}~\bibnamefont {Aolita}},\ }\bibfield  {title}
  {\bibinfo {title} {Fidelity witnesses for fermionic quantum simulations},\
  }\href {https://doi.org/10.1103/PhysRevLett.120.190501} {\bibfield  {journal}
  {\bibinfo  {journal} {Phys. Rev. Lett.}\ }\textbf {\bibinfo {volume} {120}},\
  \bibinfo {pages} {190501} (\bibinfo {year} {2018})}\BibitemShut {NoStop}%
\bibitem [{\citenamefont {Guo}\ and\ \citenamefont {Yang}(2023)}]{Guo2023A}%
  \BibitemOpen
  \bibfield  {author} {\bibinfo {author} {\bibfnamefont {Y.}~\bibnamefont
  {Guo}}\ and\ \bibinfo {author} {\bibfnamefont {S.}~\bibnamefont {Yang}},\
  }\bibfield  {title} {\bibinfo {title} {Noise effects on purity and quantum
  entanglement in terms of physical implementability},\ }\href
  {https://doi.org/10.1038/s41534-023-00680-1} {\bibfield  {journal} {\bibinfo
  {journal} {npj Quantum Inform.}\ }\textbf {\bibinfo {volume} {9}},\ \bibinfo
  {pages} {11} (\bibinfo {year} {2023})}\BibitemShut {NoStop}%
\bibitem [{\citenamefont {Endo}\ \emph {et~al.}(2021)\citenamefont {Endo},
  \citenamefont {Cai}, \citenamefont {Benjamin},\ and\ \citenamefont
  {Yuan}}]{Endo2021}%
  \BibitemOpen
  \bibfield  {author} {\bibinfo {author} {\bibfnamefont {S.}~\bibnamefont
  {Endo}}, \bibinfo {author} {\bibfnamefont {Z.}~\bibnamefont {Cai}}, \bibinfo
  {author} {\bibfnamefont {S.~C.}\ \bibnamefont {Benjamin}},\ and\ \bibinfo
  {author} {\bibfnamefont {X.}~\bibnamefont {Yuan}},\ }\bibfield  {title}
  {\bibinfo {title} {Hybrid quantum-classical algorithms and quantum error
  mitigation},\ }\href {https://doi.org/10.7566/jpsj.90.032001} {\bibfield
  {journal} {\bibinfo  {journal} {J. Phys. Soc. Jpn.}\ }\textbf {\bibinfo
  {volume} {90}},\ \bibinfo {pages} {032001} (\bibinfo {year}
  {2021})}\BibitemShut {NoStop}%
\bibitem [{\citenamefont {Cai}\ \emph {et~al.}(2008)\citenamefont {Cai},
  \citenamefont {Candes},\ and\ \citenamefont {Shen}}]{Cai2008}%
  \BibitemOpen
  \bibfield  {author} {\bibinfo {author} {\bibfnamefont {J.-F.}\ \bibnamefont
  {Cai}}, \bibinfo {author} {\bibfnamefont {E.~J.}\ \bibnamefont {Candes}},\
  and\ \bibinfo {author} {\bibfnamefont {Z.}~\bibnamefont {Shen}},\ }\href@noop
  {} {\bibinfo {title} {A singular value thresholding algorithm for matrix
  completion}} (\bibinfo {year} {2008}),\ \Eprint
  {https://arxiv.org/abs/0810.3286} {arXiv:0810.3286} \BibitemShut {NoStop}%
\end{thebibliography}%

\section{Acknowledgements}
This work is supported by the National Natural Science Foundation of China (NSFC) (Grant No. 12174214 and No. 92065205), the National Key R\&D Program of China (Grant No. 2018YFA0306504), the Innovation Program for Quantum Science and Technology (Grant No. 2021ZD0302100), and the Tsinghua University Initiative Scientific Research Program.
The IBM Quantum device `ibm\_nairobi' is accessible at \href{https://quantum-computing.ibm.com}{https://quantum-computing.ibm.com/}.
The Quafu Quantum device `ScQ-P10' is accessible at \href{https://quafu.baqis.ac.cn/}{https://quafu.baqis.ac.cn/}.

\section{Author contributions}
Y.G. conceived, designed, and performed the experiments.
Y.G. and S.Y. analyzed the data and wrote the paper.
S.Y. contributed analysis tools and supervised the project.

\section{Competing interests}
The authors declare no competing interests.

\end{document}